\documentclass[Afour,sageh,times]{sagej}

\usepackage{moreverb,url}

\usepackage[colorlinks,bookmarksopen,bookmarksnumbered,citecolor=red,urlcolor=red]{hyperref}
\usepackage{xcolor}
\usepackage{soul}
\usepackage{pgf}
\usepackage{tikz}
\usepackage{listings}
\usepackage{float}
\usepackage{multicol}
\usetikzlibrary{snakes}

\newcommand\BibTeX{{\rmfamily B\kern-.05em \textsc{i\kern-.025em b}\kern-.08em
T\kern-.1667em\lower.7ex\hbox{E}\kern-.125emX}}

\newcommand{\revision}[1]{\textcolor{black}{#1}}

\newenvironment{filecode-0}[1][]
{\minipage{\linewidth}
\lstset{framexleftmargin=0mm,basicstyle=\ttfamily\fontfamily{pcr}\footnotesize,frame=single,
#1}}
{\endminipage}

\begin{document}

\runninghead{Co-design Center for Particle Applications (CoPA)}

\title{Enabling Particle Applications for Exascale Computing Platforms}

\author{Susan M Mniszewski\affilnum{2}, 
James Belak\affilnum{3},
Jean-Luc Fattebert\affilnum{4},
Christian FA Negre\affilnum{2},
Stuart R Slattery\affilnum{4},
Adetokunbo A Adedoyin\affilnum{2},
Robert F Bird\affilnum{2}, 
CS Chang\affilnum{5},
Guangye Chen\affilnum{2},
St\'ephane Ethier\affilnum{5},
Shane Fogerty\affilnum{2},
Salman Habib\affilnum{1},
Christoph Junghans\affilnum{2},
Damien Lebrun-Grandi\'e\affilnum{4},
Jamaludin Mohd-Yusof\affilnum{2},
Stan G Moore\affilnum{6},
Daniel Osei-Kuffuor\affilnum{3},
Steven J Plimpton\affilnum{6},
Adrian Pope\affilnum{1},
Samuel Temple Reeve\affilnum{4},
Lee Ricketson\affilnum{3},
Aaron Scheinberg\affilnum{7},
Amil Y Sharma\affilnum{5},
Michael E Wall\affilnum{2}
}

\affiliation{\affilnum{1}Argonne National Laboratory, USA\\
\affilnum{2}Los Alamos National Laboratory, USA\\
\affilnum{3}Lawrence Livermore National Laboratory, USA\\
\affilnum{4}Oak Ridge National Laboratory, USA\\
\affilnum{5}Princeton Plasma Physics Laboratory, USA\\
\affilnum{6}Sandia National Laboratories, USA\\
\affilnum{7}Jubilee Development, USA}

\corrauth{S M Mniszewski, \\
Los Alamos National Laboratory, \\
P. O. Box 1663, MS B214, \\
Los Alamos, NM, USA
}

\email{smm@lanl.gov}

\begin{abstract}


The Exascale Computing Project (ECP) is invested in
co-design to assure key applications are ready for exascale computing. Within ECP, the Co-design Center for Particle Applications (CoPA) is addressing challenges faced by 
particle-based applications
across four ``sub-motifs'': short-range particle-particle 
interactions (e.g., those which often dominate molecular dynamics (MD) and smoothed 
particle hydrodynamics (SPH) methods), long-range particle-particle interactions (e.g., electrostatic  MD and gravitational N-body), particle-in-cell (PIC) methods, and 
linear-scaling electronic structure and quantum molecular dynamics (QMD) algorithms. 
Our crosscutting co-designed technologies fall into two categories: proxy applications 
(or “apps”) and libraries. Proxy apps are vehicles used to evaluate the viability of incorporating various types of algorithms, data structures, and architecture specific optimizations, and the associated trade-offs; examples include ExaMiniMD, CabanaMD, CabanaPIC, and ExaSP2. 
Libraries are modular instantiations that multiple applications can utilize or be built upon; CoPA has developed the Cabana particle library, PROGRESS/BML libraries for QMD, and the SWFFT and fftMPI parallel FFT libraries.
Success is measured by identifiable “lessons learned” that are translated either 
directly into parent production application codes or into libraries, with 
demonstrated performance and/or productivity improvement. 
The libraries and their use in CoPA's ECP application partner codes are also addressed.
\end{abstract}


\maketitle


\section{Introduction}
The US DOE Exascale Computing Project (ECP) Co-design Center for Particle Applications (CoPA) provides contributions to enable application readiness as we move toward exascale architectures for the ``motif''  of particle-based applications \cite{ecp}.
CoPA focuses on co-design for the following ``sub-motifs'': short-range particle-particle interactions (e.g., those which often dominate molecular dynamics (MD) and smoothed particle hydrodynamics (SPH) methods), long-range particle-particle interactions (e.g., electrostatic  MD and gravitational N-body), particle-in-cell (PIC) methods, and 
O(N) complexity electronic structure and quantum molecular dynamics (QMD) algorithms. 

\begin{figure*}
\includegraphics[width=16cm]{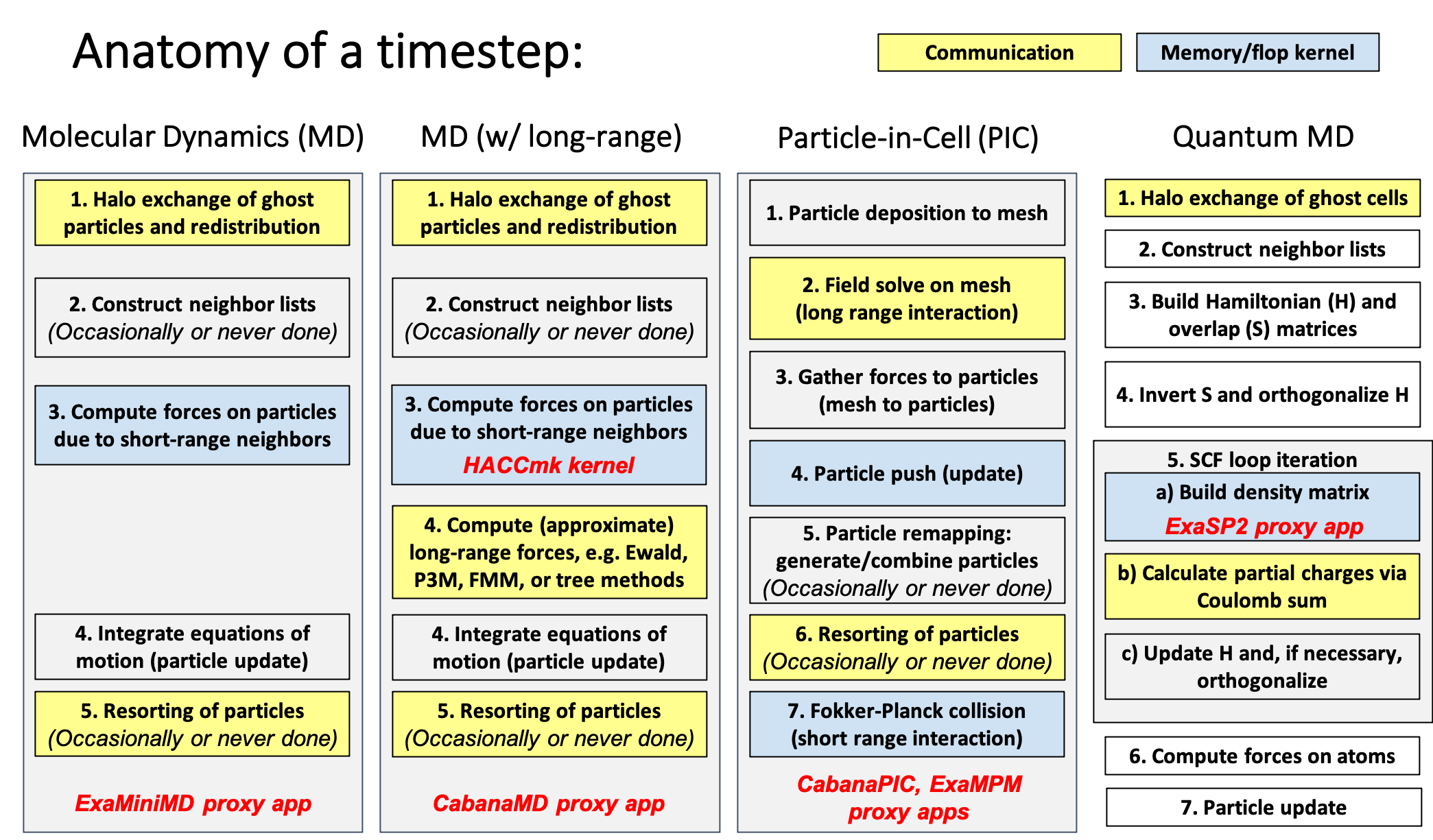}
\caption{The anatomy of a timestep is shown for each of the particle application \revision{sub-motifs} addressed in CoPA. Communication intensive steps and compute/memory intensive steps are shown in yellow and blue respectively.}
\label{intro:anatomyTS}
\end{figure*}

Particle-based simulations start with a description of the problem in terms of particles and commonly use a single-program multiple-data domain decomposition paradigm for inter-node parallelism.
\revision{Because particles in each domain interact with particles outside its domain, a list of these outside particles must be maintained and updated through inter-node communication.
This list of outside particles is commonly kept in a set of ghost cells or a ghost region on each node.}
Intra-node parallelism is commonly performed through work decomposition.
From a description of the neighborhood (neighbor list), each particle's forces are calculated to propagate the particles to new positions.
The particles are then resorted to begin the next timestep.
The main components of a timestep across the sub-motifs are shown in Figure~\ref{intro:anatomyTS}.
PIC is unique in that particles are used to solve continuum field problems on a \revision{grid}.
QMD solves the computationally intensive electronic-structure for problems where details of inter-atomic bonding are particularly important. Shared and specific functionality are highlighted in Figure~\ref{intro:anatomyTS}.
The compute, memory, and/or communication challenges requiring optimization on modern computer architectures are identified, extracted and assembled into libraries and proxy applications \revision{during the CoPA co-design process}.


CoPA's co-design process of using proxy applications (or apps) and libraries
has grown out of \revision{a predecessor project,} the Exascale Co-Design Center for Materials in Extreme Environments (ExMatEx) \cite{exmatex}. Two main library directions have emerged, the Cabana Particle Simulation Toolkit and the PROGRESS/BML QMD Libraries, each described in later sections.
Each strive for performance portability, flexibility, and scalability across architectures with and without GPU acceleration by providing optimized data structure, data layout, and data movement in the context of the sub-motifs they address. Cabana is focused on short-range and long-range particle interactions for MD, PIC, and N-body applications, while PROGRESS/BML is focused on O(N) complexity algorithms for electronic structure and QMD applications. \revision{This split is primarily motivated by the difference in sub-motifs: QMD is computationally dominated by matrix operations, while the 
other sub-motifs share particle and particle-grid operations.}
The locations for the open-source CoPA libraries and \revision{proxy apps} are noted in the sections in which they are described.


\revision{The particle motif is used by many application codes to describe physical systems, including molecular dynamics simulations using empirical models or the underlying quantum mechanics for particle interactions, cosmological simulations in which the particle may represent an object (e.g. a star) or a cluster of objects and the particle interaction is through gravity, and plasma simulations 
on grids within a PIC framework to solve the interaction of particles with the electro-dynamic field. The computational motifs associated with these application codes depends on the nature of the particle interactions. Short-ranged interactions rely heavily on the creation of a list of neighbors for direct interactions, while long-range interactions use particle-grid methods and FFTs to solve the long-range field problem. Details are described in the section on the Cabana Toolkit. The quantum mechanics in QMD problems is often expressed as a matrix problem. 
QMD based on localized orbitals in Density Functional Theory (DFT) and tight-binding models are reliant on sparse-matrix solvers. Details are described in the section on the PROGRESS/BML Libraries.}

\begin{figure}[h]
\includegraphics[width=8cm]{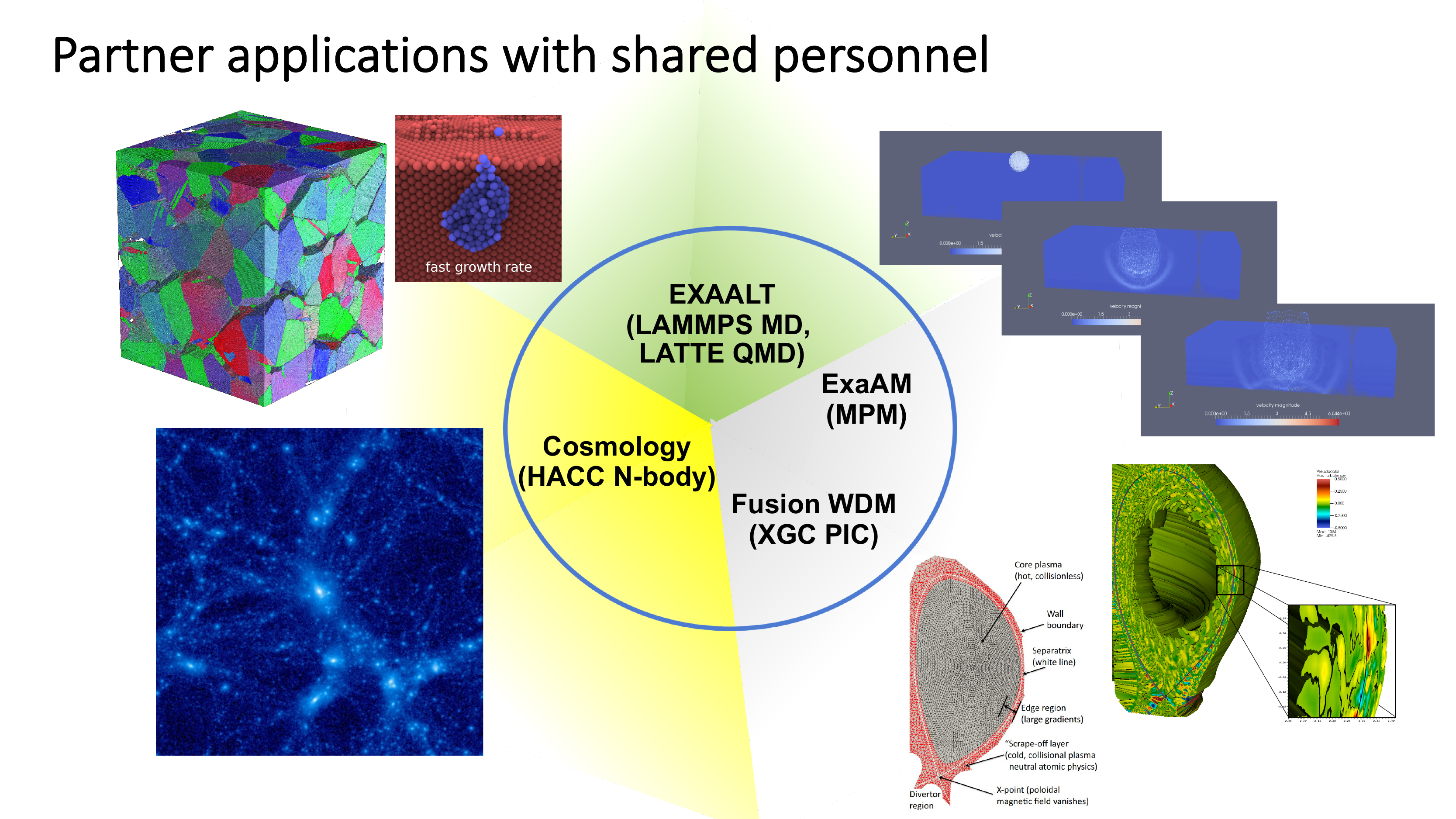}
\caption{Partner ``particle motif'' applications with shared personnel are shown. \revision{These applications represent all the CoPA sub-motifs.}}
\label{intro:partnerapps}
\end{figure}

Relevant particle applications are represented within CoPA and help drive the co-design process. ECP application projects such as EXAALT (LAMMPS-SNAP), WDMApp (XGC), ExaSky (HACC/SWFFT), and ExaAM (MPM) serve as application partners as shown in Figure~\ref{intro:partnerapps}, as well as non-ECP applications. \revision{Details of these engagements are described in the section on Application Partners.}

We present descriptions of the Cabana Particle Simulation Toolkit and PROGRESS/BML QMD libraries, followed by PIC algorithm development, and co-design examples with our application partners: XGC, HACC, and LAMMPS-SNAP. We conclude with a summary of our lessons learned and impact on the broader community.


\section{Cabana Particle Simulation Toolkit}

The Cabana toolkit is a collection of libraries and proxy applications which allows scientific software developers targeting exascale machines to develop scalable and portable particle-based algorithms and applications. The toolkit is an open-source implementation of numerous particle-based algorithms and data structures applicable to a range of application types including (but not limited to) PIC and its derivatives, MD, SPH, and N-body codes \cite{hockney_eastwood,liu2003smoothed} and is usable by application codes written in C++, C, and FORTRAN. \revision{Notably, this covers the first three sub-motifs (see Figure \ref{intro:anatomyTS}). Cabana is designed as a library particularly because so many computational algorithms are shared across particle applications in these sub-motifs: neighbor list construction, particle sorting, multi-node particle redistribution and halo exchange, etc. This effectively separates shared capabilities from the specific application physics within individual steps of Fig. \ref{intro:anatomyTS}, e.g. the per-atom force computation in MD and the particle-grid interpolation for PIC.}
Cabana is available at \url{https://github.com/ECP-CoPA/Cabana}.

The toolkit provides both particle algorithm implementations and user-configurable particle data structures. Users of Cabana can leverage the algorithms and computational kernels provided by the toolkit independent of whether or not they are also utilizing the native data structures of the toolkit through memory-wrapping interfaces. The algorithms themselves span the space of particle operations necessary to support each relevant application type\revision{, spanning across all sub-motifs}. This includes intra-node (local and threaded) operations on particles and inter-node (communication between nodes) operations to form a hybrid parallel capability. Cabana uses the Kokkos programming model for on-node parallelism \cite{kokkos}, providing performance and portability on pre-exascale and anticipated exascale systems using current and future DOE-deployed architectures, including multi-core CPUs and GPUs. Within Cabana, Kokkos is used for abstractions to memory allocation, array-like data structures, and parallel loop concepts which allow a single code to be written for multiple architectures.

\begin{figure}[h]
\includegraphics[width=8cm]{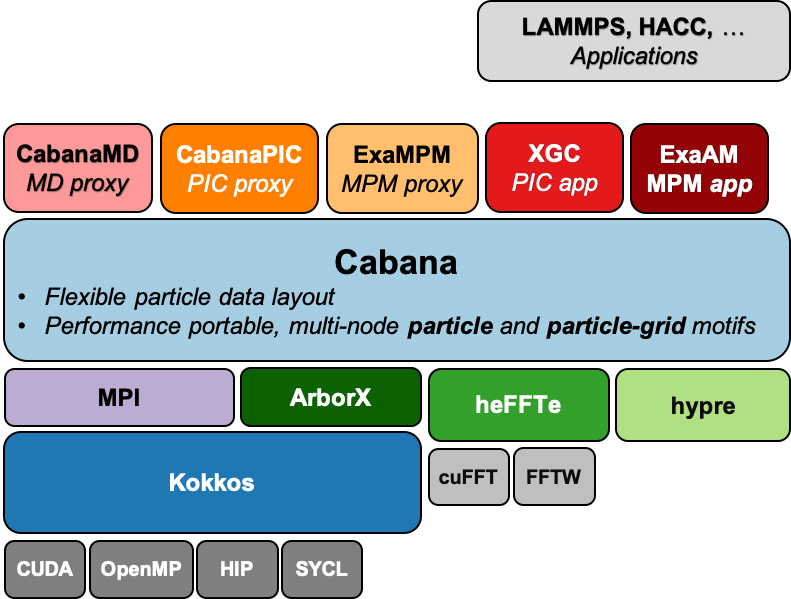}
\caption{Cabana software stack including dependencies, proxy apps, and production apps.}
\label{cabana:stack}
\end{figure}

While the only required dependency for Cabana is Kokkos, the toolkit is also intended to be interoperable with other ECP scientific computing libraries which the user may leverage for functionality not within the scope of the toolkit. Use of the library in concert with other ECP libraries can greatly facilitate the composition of scalable particle-based application codes on new architectures. Current library dependencies are shown in Figure~\ref{cabana:stack} \revision{with} libraries developed by the ECP \revision{Software Technology (ST)} projects,
including hypre for preconditioners and linear solvers \cite{hypre}, heFFTe for 3D-distributed FFTs \cite{heffte}, and ArborX \cite{arborx} for threaded and distributed search algorithms.

\revision{Cabana includes both particle and particle-grid operations which are critical across the particle sub-motifs. We next review each of these capabilities.}

\subsection{Particle Abstractions}

\revision{We first} summarize the major particle-centric abstractions and functionality of the toolkit including the underlying data structure and the common operations which can be applied to the particles.

\subsubsection{Data Structures}

\begin{figure}[]
\includegraphics[width=8cm]{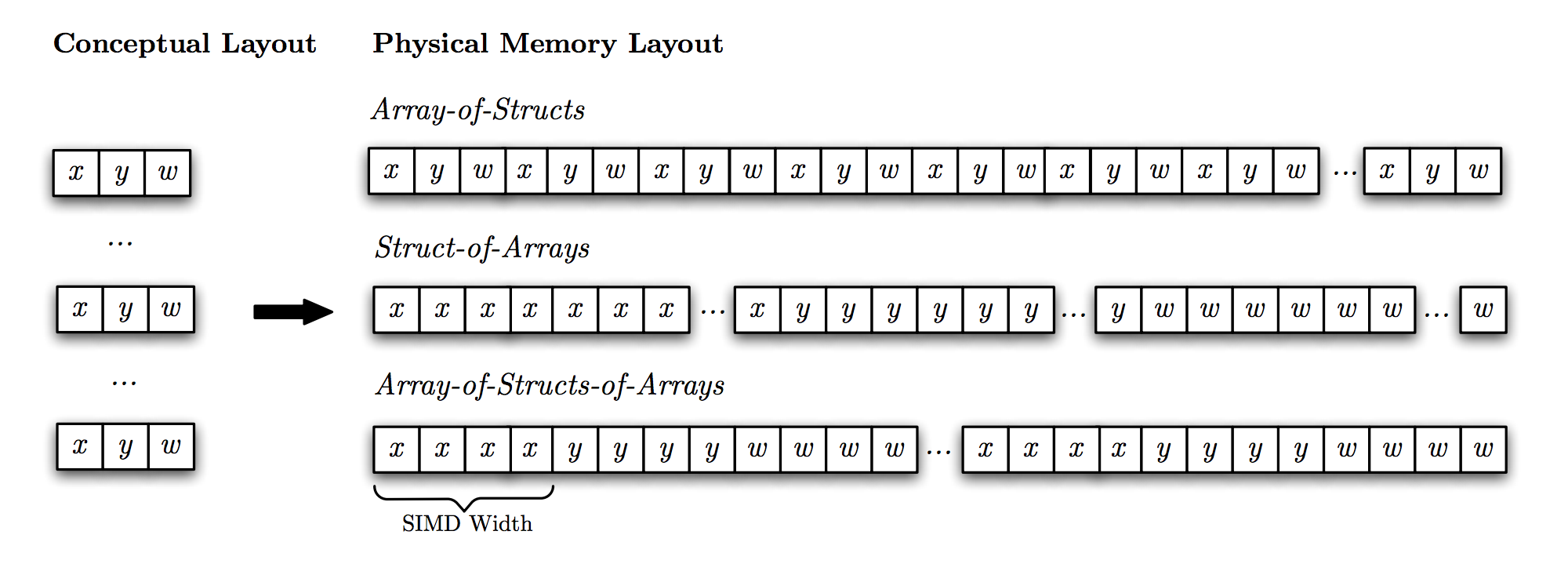}
\caption{Particles in Cabana are stored in an Array-of-Structs-of-Arrays (AoSoA). Compared to an Array-of-Structs (AoS), an AoSoA will provide similar memory locality benefits during data access while also coalescing data access over SIMD-length elements when possible. Compared to a Struct-of-Arrays (SoA), an AoSoA will provide similar coalescing data access benefits while also introducing additional memory locality.}
\label{cabana:aosoa}
\end{figure}

Particles in Cabana are represented as tuples of multidimensional data. They are general and may be composed of fields of arbitrary types and dimensions (e.g. a mass scalar, a velocity vector, a stress tensor, and an integer id number) as defined by the user's application. 
Considering the tuple to be the fundamental particle data structure (struct), several choices exist for composing groups of structs as shown in Figure \ref{cabana:aosoa}. 
A simple list of structs, called an Array-of-Structs (AoS) is a traditional choice for particle applications, especially those not targeting optimization for vector hardware. 
All of the data for a single particle is encapsulated in a contiguous chunk of memory, thereby improving memory locality for multiple fields within a particle. 
An AoS also offers simplicity for basic particle operations, such as sorting or communication, as \revision{the memory associated with a given particle may be manipulated in a single operation.}
An AoS, however, also has a downside: accessing \revision{the same data component in} multiple particles concurrently requires strided (non-coalesced) memory accesses which incurs a significant performance penalty on modern vector-based machines.

This penalty for strided access can be alleviated through the use of the Struct-of-Arrays (SoA) memory layout. In an SoA, each particle field is stored in a separate array, with the values of an individual field stored contiguously for all particles. This structure allows for a high-performance memory access pattern that maps well to modern vector-based architectures. The drawback of this approach, however, is two-fold: first, the hardware has to track a memory stream for each particle property used within a kernel and second, the programmer and hardware may have a harder time efficiently operating on all of the data together for a given particle.    
In light of these features, it is typically favorable to use SoA when effective use of vector-like hardware is vital (as is the case with GPUs), or when a subset of particle fields are used in a given kernel. 

Cabana offers a zero-cost abstraction to these memory layouts, and further implements a hybrid scheme known as Array-of-Structs-of-Arrays (AoSoA). An AoSoA attempts to combine the benefits of both AoS and SoA by offering user-configurable sized blocks of contiguous particle fields. This approach means a single memory load can fetch a coalesced group of particle field data from memory, while retaining high memory locality for all fields of a given particle.
A key performance tuning parameter exposed by Cabana, the added AoSoA dimension enables the user to configure the memory layout of a given particle array at compile time or to have it automatically selected based on the target hardware.

\subsubsection{Particle Sorting}
Particle sorting is a functionality requirement of all currently identified user applications. In plasma PIC, fluid/solid PIC, MD, N-body, and SPH calculations, particle sorting serves as a means of improving memory access patterns based on some criteria which can provide an improvement in on-node performance. This could be, for example, placing particles that access the same grid cell data near each other in memory, or grouping particles together by material type such that particles adjacent in memory will be operated on by the same computational kernel in the application. The frequency of sorting often depends on the application, as well as the given problem. Slowly evolving problems may need to sort particles less frequently as local particle properties that affect memory access (e.g. grid location or nearest neighbors) will be relatively stable. Efficient memory access objectives should be based on the target computing platform (e.g. GPUs or multi-CPU devices). Particle sorting is applicable to locally owned particles or to both locally owned and ghost particles. Cabana provides the ability to sort particles by spatial location through geometric binning or by an arbitrary key value, which can include either a particle property or another user-provided value. \revision{Cabana uses the bin sort functionality in Kokkos, with plans for additional options through Kokkos in the future.}

\subsubsection{Neighbor List Creation}
PIC, MD, N-body, and SPH can all benefit from the efficient construction of particle neighbor lists. These neighbor lists are typically generated based on a distance criteria where a physical neighborhood is defined or instead based on some fixed number of nearest neighbors. In both MD and SPH simulations the neighbor list is a critical data structure and is computed more frequently than most other particle operations up to the frequency of every time step. In fluid/solid PIC applications the neighbor list is an auxiliary data structure for computational convenience that, when used, is similarly computed up to the frequency of every time step. In the fluid/solid PIC case the background grid can be used to accelerate the neighbor search, whereas in MD, N-body, and SPH a grid may need to be created specifically for this search acceleration. Cabana provides multiple variations of neighbor lists, including traditional binning methods used in many MD applications (Verlet lists), as well as tree-based algorithms from the ECP ST ArborX library \cite{arborx}, using compressed and dense storage formats, and thread-parallel hierarchical list creation; all of these options can improve performance for different architectures and particle distributions.

\subsubsection{Halo Exchange and Redistribution}
Many user applications require parallel communication of particles between compute nodes when spatial domain decomposition is used and particle data must be shared between adjacent domains. Some applications require halo exchange operations on particles and some, in particular \revision{grid}-free methods (e.g. MD), additionally require ghost particle representations to complete local computations near domain boundaries. In many cases, the halo exchange is executed at every single time step of the simulation with a communication pattern that may also be computed at every time step or at some larger interval and reused between constructions. In addition, particles need to be redistributed to new compute nodes in many algorithms either as a result of a load balancing operation or because advection has moved particles to a new region of space owned by another compute node. The toolkit provides implementations for ghost particle generation and halo exchange, including both gather and scatter operations, as well as a migration operation to redistribute particles to new owning domains.

\subsubsection{Parallel \revision{Loops}}
Cabana adds two main extensions to the Kokkos parallel constructs which handle portable threaded parallelism and mapping hierarchical and nested parallelism to up to three levels on the hardware. First, SIMD parallel loops are directly connected to the AoSoA data structures and provide the user simple iteration over the added inner vector dimension, for threaded parallelism over both particle structs and vector, with potential performance improvements \revision{by exposing coalesced memory operations}. Second, neighbor parallel loops provide both convenience and flexibility for any particle codes which use a neighbor list (see above) to iterate over both particles and neighboring particles (including first and/or second-level neighbors). Cabana handles all neighbor indexing and the user kernel deals only with application physics. This also enables applications to easily change the parallel execution policy, and use the appropriate threading \revision{over the neighbor list structures} (or serial execution) for the kernel, problem, and hardware. 

\subsection{Particle-Grid Abstractions}

In addition to pure particle abstractions, the toolkit also contains optional infrastructure for particle-grid concepts which we present next.

\subsubsection{Long Range Solvers}
Among the user applications, long range solvers encapsulate a wide variety of kernels and capabilities, but many include critical kernels that can apply to many applications. For example, embedded within the long range solve of a PIC operation are kernels for interpolating data from the particles to the grid and from the grid to the particles, as well as possibly grid-based linear solvers. Other simulations, such as MD and SPH calculations, compose the long range solvers with particle-grid operations, but instead use other algorithms such as Fast Fourier Transforms (FFT) to complete the long range component of the solve.  A variety of libraries provide FFT capabilities, including high-performance scalable FFT libraries being developed for large systems. Cabana currently implements direct use of the ECP ST heFFTe library for performance portable FFTs \cite{heffte} and Cabana's flexibility will enable use of other ECP FFT libraries in the future, including FFTX \cite{fftx}, SWFFT \cite{swfft}, and fftMPI \cite{fftmpi}. \revision{In addition, Cabana interfaces to hypre \cite{hypre} to provide interfaces to linear solvers.}

\subsubsection{Particle-Grid Interpolation}
PIC methods, as well as methods which require long range solvers, usually need some type of interpolation between particle and grid representations of a field \revision{in order to populate the grid data needed by FFTs, linear solvers, or other field operations}. The \revision{toolkit provides services for} interpolation to logically structured grids based on multidimensional spline functions, which are available in multiple orders. By differentiating the spline functions, differential operators may be composed during the interpolation process, allowing users to interpolate gradients, divergences, and other operators of scalar, vector, and tensor fields. \revision{Other types of interpolants can also be added in the future for additional capabilities in PIC applications. As needed, we also envision generalizing the Cabana interpolation infrastructure to support user-defined interpolants on structured grids.}

\subsection{Proxy apps}
\revision{Cabana-based} proxy apps have been developed \revision{in order} to demonstrate and improve Cabana functionality as well as to explore new algorithms and ideas \revision{when they are deployed in the context of a sub-motif. In addition to those presented here, more proxies are planned to cover additional variations of the algorithmic abstractions represented by application partners}.

\subsubsection{CabanaMD}
\begin{figure}[h]
\includegraphics[width=8cm]{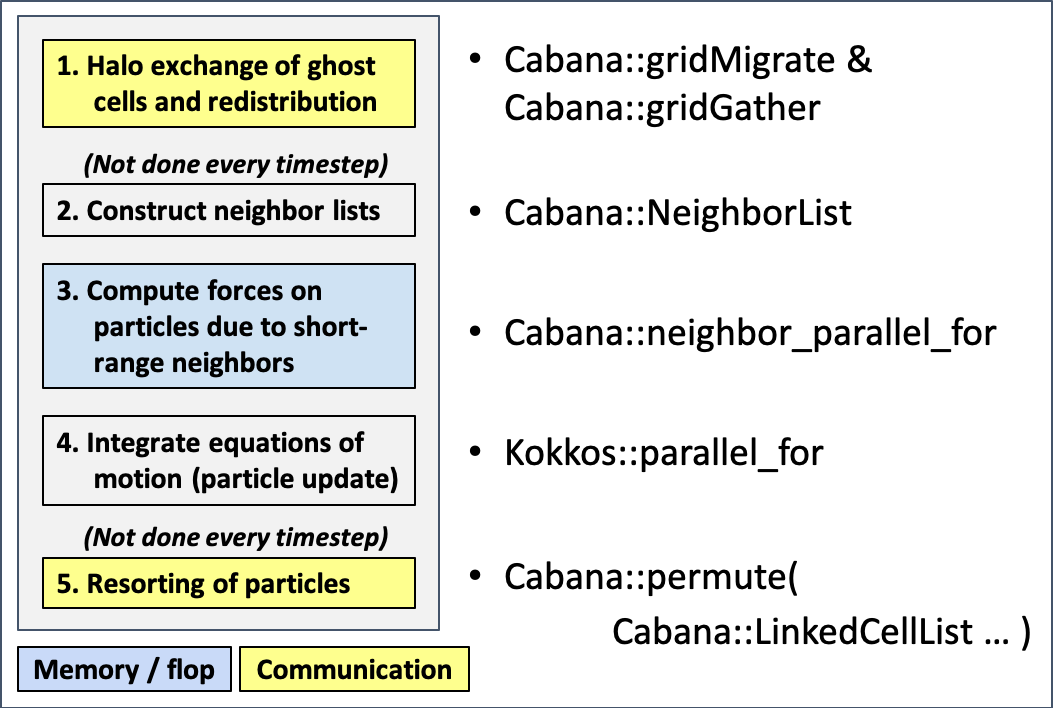}
\caption{Single MD timestep re-expressed with the Cabana/Kokkos API. We have purposefully mapped our API to the main algorithmic components of a timestep.}
\label{cabana:step}
\end{figure}

CabanaMD is a LAMMPS \cite{Plimpton1995} proxy app built on Cabana, developed directly from the ExaMiniMD proxy (\revision{``}KokkosMD") \cite{kokkos}. \revision{CabanaMD represents both the short and long range MD sub-motifs}; in fact, the MD timestep can easily be re-expressed as calls to the Cabana library, as shown in Figure \ref{cabana:step}. Similar figures could also be created for all sub-motifs in Figure \ref{intro:anatomyTS} with all the CoPA proxy apps.

CabanaMD is available at \url{https://github.com/ECP-CoPA/CabanaMD}. MD uses Newton's equations for the motion of atoms, with various models for interatomic forces and ignoring electrons. In contrast to ExaMiniMD and many applications, only the main physics, the force kernel evaluating the interatomic model and position integration, is entirely retained in the application with everything else handled by Cabana. Main results for demonstrating Cabana capabilities include performance implications of combining or separating particle properties in memory, changing the particle property AoSoA memory layouts, and using the available options for each algorithm (e.g. data layout, hierarchical creation, and hierarchical traversal for neighbor lists). This has been done primarily with the Lennard-Jones short range force benchmark kernel. In addition, CabanaMD enabled \revision{speedup on the order of 3x (one Nvidia V100 GPU compared to a full IBM Power 9 CPU node)} in new neural network interatomic models \cite{behler} by re-implementing with Kokkos and Cabana, rewriting the short range force kernels, and exposing threaded parallelism \cite{desai}.

CabanaMD also includes long range forces using Cabana data structures and particle-\revision{grid} algorithms\revision{, covering the second sub-motif}. The smooth particle mesh Ewald (SPME) \cite{spme} method is implemented with Cabana \revision{grid} structures and spline kernels to spread particle charge onto a uniform grid, 3D FFTs using the Cabana interface to heFFTe \cite{heffte} for reciprocal space energies and forces, and Cabana gradient kernel to gather force contributions back from the \revision{grid} to atoms. Real-space energy and force calculations use the Cabana neighbor parallel iteration options, as with short range interactions. \revision{Continuing work for CabanaMD will include benchmarking long range performance and using it as a vehicle to implement and improve performance portable machine learned interatomic models.}

\subsubsection{CabanaPIC}

CabanaPIC is a relativistic PIC proxy app using Cabana, capable of modeling kinetic plasma \revision{and representing the PIC sub-motif. CabanaPIC is} available at \url{https://github.com/ECP-CoPA/CabanaPIC}. It has strong ties with the production code VPIC \cite{vpic}, but is able to act as a representative proxy for all traditional electromagnetic PIC codes which use a structured grid. It implements a typical Boris pusher, 
as well as a finite-difference time-domain (FDTD) field solver.

CabanaPIC focuses on short range particle-\revision{grid} interactions, and its performance is heavily dependent on techniques to martial conflicting writes to memory (such as atomics). It employs Cabana's particle sorting techniques, as well as offers examples of how to use Cabana for simple MPI based particle passing.

\subsubsection{ExaMPM}


ExaMPM is a Cabana-based proxy application for the Material Point Method (MPM) which is being used as part of high-fidelity simulations of additive manufacturing with metals in the ExaAM project in ECP. 
ExaMPM \revision{also represents the PIC sub-motif and} is available at \url{https://github.com/ECP-CoPA/ExaMPM}. MPM, a derivative of PIC, is used to solve the Cauchy stress form of the Navier-Stokes equations including terms for mass, momentum, and energy transport where particles track the full description of the material being modeled in a Lagrangian and continuum sense. The MPM simulations in ExaMPM model the interaction of a laser with metal powders, the melting of the powder due to heating from the laser, and the solidification of the melted substrate after the laser is turned off. When modeled at a very high-fidelity, using particles to track the free surface interface of the molten metal and the liquid-solid interface during phase change will allow for both empirical model generation in tandem with experiments, as well as reduced-order model generation to use with engineering-scale codes in the ExaAM project.

\begin{figure}[h]
\includegraphics[width=8.5cm]{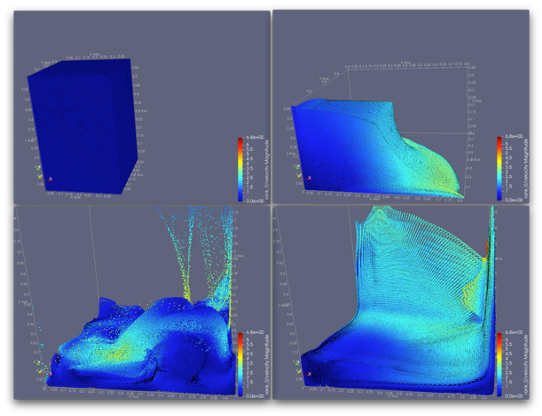}
\caption{Snapshots of a 3D water column collapse modeled with ExaMPM (ordered clockwise starting from top-left). Complex moving interfaces and dynamic local particle populations are being used to study improved algorithms for load balancing, communication, and particle sorting.}
\label{cabana:exampm}
\end{figure}

ExaMPM largely implements the base algorithmic components of the MPM model including an explicit form of time integration, a free surface formulation with complex moving interfaces, and higher-order particle-grid transfer operators which reduce dissipation. As an example, Figure~\ref{cabana:exampm} shows a water column collapse modeled with ExaMPM. This problem has a dynamic moving surface structure resembling the dynamics of the molten metal in the high-fidelity ExaAM simulations as well as particle populations which change rapidly with respect to the local domain. Scaling up this problem \revision{through larger particle counts and larger computational meshes will allow} us to study better techniques for interface tracking, more scalable communication and load balancing algorithms to handle the moving particles, and sorting routines to improve locality in particle-grid operations.

\subsection{Cabana Applications}
A significant metric for the impact of a given software library is its adoption. First, Cabana is being used as the basis for a new production application closely related to the ExaMPM proxy described in the previous section. Another notable usage of Cabana is in the transition of an existing application; the section on XGC-Cabana below details the process of converting XGC, initially using Cabana through FORTRAN interfaces, eventually to full C++ with Cabana.

In addition, the Cabana library has and will continue to influence production applications and related libraries. This includes LAMMPS and HACC (also described in later sections), where the performance of an algorithm, data layout, etc. can be demonstrated with Cabana and/or its proxy apps and migrated to the separate application; similarly, interactions between Cabana and the libraries it depends on can improve each for a given application. 

\section{PROGRESS/BML Quantum Molecular Dynamics Libraries} 

\revision{This section focuses on the solvers addressing the quantum part of Quantum Molecular Dynamics (QMD), the sub-motif listed in the fourth column of Fig.\ref{intro:anatomyTS}.}
QMD uses electronic structure (ES) based atomic forces 
to advance the position of classical particles (atoms)
in the Born-Oppenheimer approximation \cite{Marx2009-zg}. There are many benefits of this technique as compared to classical methods. These benefits include: independence of the results associated with the choice of a particular force field;
enabling the formation and breaking of bonds (chemical reactions) as the simulation proceeds; and, the possibility of extracting information from the electronic structure throughout the simulation.
The counterpart to these benefits is the large computational cost associated with having to determine the ES of the system before advancing the particles coordinates. In order to perform practical MD simulations, the strong scaling limit becomes important, as time-to-solution needs to be as small as possible to enable long simulations
with tens of thousands of timesteps, and achieve a good sampling of the phase-space.
Determining the ES is the main bottleneck of QMD, where the so called single-particle
density matrix (DM) needs to be computed from the Hamiltonian matrix. The latter requires a significant amount of arithmetic operations, and
it typically scales with the cube of the number of particles. QMD is hence characterized by this unique critical step that sets it apart from all the particle simulation methods within the scope of the Cabana toolkit (described in the previous section); therefore, additional libraries are needed for increasing its performance and portability. 

Another significant challenge surrounding the development of QMD codes is that solvers used to compute the ES are strongly dependent on the chemical systems (atom types and bonds), which implies that it is necessary to develop and maintain several different algorithms that are suitable for each particular system.
For instance, the single-particle DM for insulators (with a wide energy gap between the highest-occupied and lowest-unoccupied state), is essentially a projector onto the subspace spanned by the eigenfunctions associated with
the lowest eigenvalues of the Hamiltonian matrix 
(occupied states). For metals, on the other hand, the DM is not strictly a projector since a temperature 
dependent weight between 0 and 1 is associated with each eigenstate (Fermi-Dirac distribution).
Nevertheless, in both cases the DM can be computed from the knowledge
of the eigenpairs of the Hamiltonian which are computationally expensive to determine.

In general, we can say that the construction of the DM $\rho$ can be expressed as a matrix function of the Hamiltonian $H$. Such a function  $\rho = f(H)$, can be computed exactly by diagonalizing matrix $H$.  The function then becomes: $\rho = Cf(\epsilon)C^T$, where $\epsilon$ is a diagonal matrix containing the eigenvalues of $H$, and $C$ is a unitary transformation where its columns contain the eigenvectors of $H$. $f$ is the Fermi distribution functions for finite temperatures. 

%
%
 
\begin{figure}[h]
\includegraphics[width=8cm]{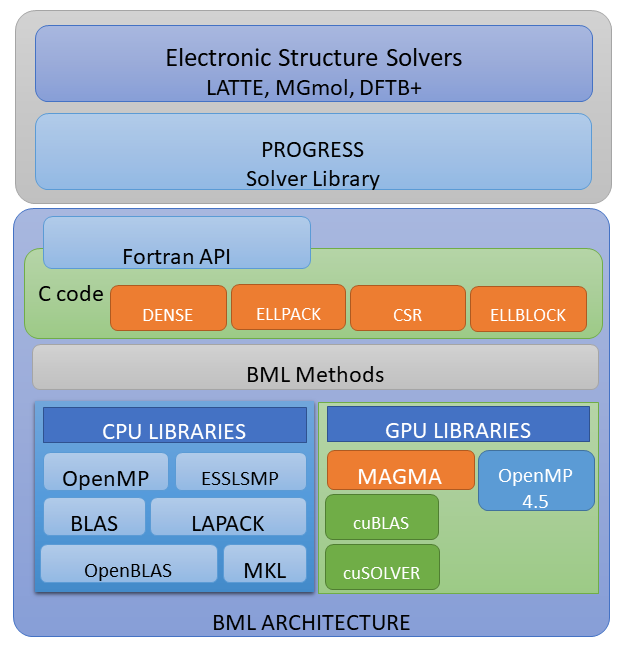}
\caption{PROGRESS/BML software stack including dependencies and production apps.}
\label{bml:stack}
\end{figure}

In order to increase productivity in the implementation and optimization of
these algorithms we have adopted a framework in which we clearly separate the
matrix operations from the solver implementations. 
The framework relies on two main libraries: ``Parallel, Rapid O(N) and Graph-based Recursive Electronic Structure Solve'' (PROGRESS), and the ``Basic Matrix Library'' (BML). The software stack can be seen in Figure \ref{bml:stack}. At the highest level, electronic structure codes 
call the solvers in the PROGRESS library which, in turn, rely on BML. 
The BML library provides basic matrix data structure and operations. These consist of linear algebra matrix operations which are optimized based
on the format of the matrix and the architecture where the program will run. 
Applications can also directly implement specific algorithms
based on BML when those are not available routines in PROGRESS.
Both libraries use travis-ci and codecov.io for continuous
integration and code coverage analysis respectively. Every commit is tested over a set of compiler and compiler options.
Our overarching goal is to construct a flexible library ecosystem that helps to quickly adapt and optimize electronic structure applications on exascale architectures. 
\revision{
Alternative libraries that overlap with the matrix formats
and algorithms implemented in PROGRESS and BML include
DBCSR \cite{dbcsr-software} and NTPoly \cite{DAWSON2018154},
both focusing also on electronic structure applications.
}

\subsection{Basic Matrix Library}

The increase in availability of heterogeneous computer platforms is the motivation behind the development of 
the BML software library.  Multiple data storage formats (both for sparse and dense) and programming models (distributed, 
threaded, and task-based) complicate the testing and optimization of electronic structure codes. 

\begin{figure}
\centering
\includegraphics[width=7.5cm]{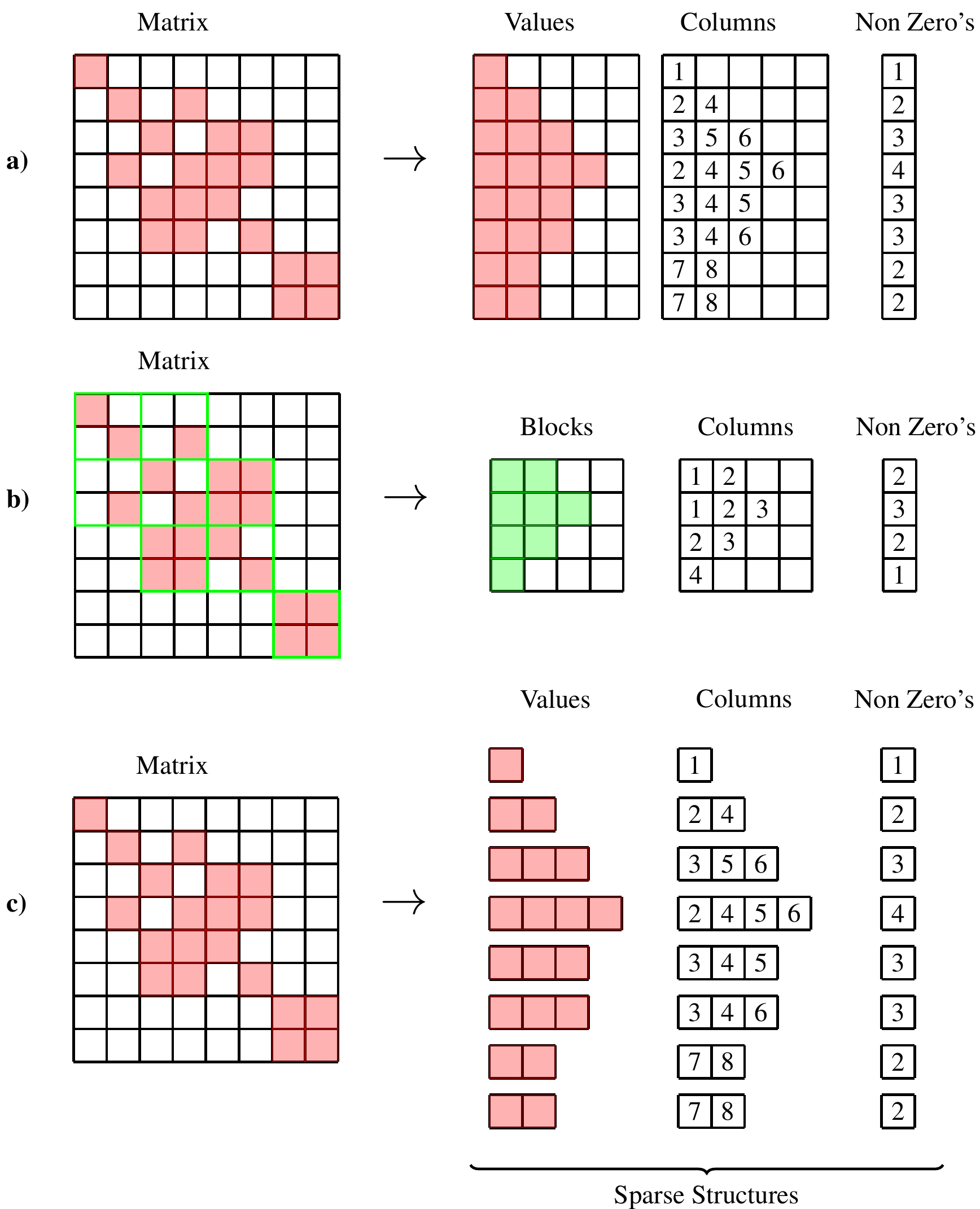}
\caption{\revision{
The three sparse matrix formats currently available in BML. a) ELLPACK-R: A 2D array containing the compressed rows; a second 2D array containing the column indices; and a third 1D array containing the maximun non-zeros per row. b) ELLBLOCK: Block version of ELLPACK-R where a matrix is decomposed into blocks that are either 
considered full of zeroes and not stored, or dense blocks that are treated as all non-zeroes.
c) CSR: 
A matrix is represented as an array of compressed sparse rows, where each compressed row stores only the nonzero entries and the associated column indices of the corresponding row of the matrix.
}}
\label{formats}
\end{figure}

The basic matrix 
library package (BML) contains the essential linear algebra operations that are relevant to electronic structure problems. The 
library is written in C, which allows for straightforward implementation on
exascale machines. 
The library also exposes a Fortran interface, with Fortran wrappers written around C functions.
This facilitates its usage by a wide variety of codes since many applications codes in this community are written in Fortran.
One of the main advantages of BML is that the APIs are the same for all matrix types, data types, and architectures, enabling users to build unified solvers that work for multiple
matrix formats.
Low-level implementations within the 
BML library are tailored to particular matrix formats and computer architectures. 
The formats that can be handled so far are: dense, ELLPACK-R \cite{Mniszewski2015}, 
\revision{Compressed Sparse Row (CSR) \cite{Saad_book}}, and ELLBLOCK
\revision{(see Figure~\ref{formats})}. 
Here, dense is used to refer to a typical two-dimensional array. It is the most suitable format for treating systems that have a high proportion of non-zero values in the DM. 
ELLPACK-R is a sparse matrix format constructed using three arrays: a one-dimensional array used to keep track of the number of non-zeros per row for each row; a two-dimensional array used to keep track of the column indices of the non-zero values within a row; and finally, another two-dimensional array used to store the nonzero values. The row data are zero-padded to constant length, so row data are separated by constant stride. 
ELLBLOCK is a block version of ELLPACK-R. 
In a nutshell, a matrix is decomposed into blocks that are either 
considered full of zeroes and not stored, or dense blocks that are treated as all non-zeroes.
Loops over nonzero matrix elements are replaced by loops over
nonzero blocks,
and dense linear algebra operations are done on blocks.
The CSR format in its typical implementation utilizes 
a floating point array to store the nonzero entries of the matrix in row-major order and an integer array to store the corresponding column indices. 
In addition, an array which indexes the beginning of each row in the data arrays is needed for accessing the data. 
Unlike the ELLPACK-R format, which stores entries in a two-dimensional array with a fixed width for all the nonzero entries in the rows of the matrix, the CSR format keeps the variability in the number of non-zeros per row, thus avoiding the need for zero-padding. The implementation of the CSR format in 
BML follows an \revision{Array-of-Structs-of-Arrays (AoSoA)} approach. 
A matrix is represented as an array of compressed sparse rows, where each compressed row stores only the nonzero entries and the associated column indices of the corresponding row of the matrix. 
In this approach, the additional array of indexes to the beginning of each row is no longer needed. 
The matrix stored in this way is extensible, allowing the matrix to grow by simply adding new entries without the need to destroy the matrix.

BML dense matrix functions are typically wrappers on top of a vendor
optimized library, such as BLAS/LAPACK implementations.
For NVIDIA GPUs, we use the MAGMA library for dense matrices \cite{dghklty14}.
We use its memory allocation functions and many of its functionalities.
One exception is the dense eigensolver for which we use the NVIDIA cuSOLVER which performs substantially better than the MAGMA solver
at the moment.
In the case of the sparse formats, each BML function is specifically written for that particular format.
Performance portability is achieved by keeping one codebase with a high flexibility for configuring and building. BML was compiled and tested with multiple compilers (GNU, IBM, Intel, etc.) on several pre-exascale architectures. 

BML ELLPACK-R matrix functions are implemented with OpenMP, both on CPU and GPU, the latter using target offload. 
The algorithm implemented utilizes a work array of size $(N)$ per row which is larger than GPU cache for matrix sizes of interest, leading to poor performance on GPU. 
Previous work by \cite{JamalGTC2015} demonstrated the performance of a novel merge-based implementation of sparse-matrix multiply on GPU, implemented in CUDA. 
Future implementations will utilize a mix of OpenMP offload and native CUDA kernels to enable performance while retaining a consistent interface with the existing OpenMP implementations. 
Benchmarking indicates this should allow a speedup of \revision{${\sim}8\times$ on an} Nvidia V100 compared to IBM Power 9
\revision{(using all 21 cores of one socket)}. 

BML offers support for four datatypes: 
single precision, double precision, complex single precision, and complex double precision.
The source code for all these datatypes is the same
for most functions, with C macros that are preprocessed at compile time to generate functions for the four
different formats.
All BML function names are prefixed with \verb|bml_|. 
The code listing in Figure~\ref{sp2} shows the use of the BML API on one of our O(N) complexity algorithms, the ``second order spectral projection'' (SP2)
\cite{NiklassonTymczakChallacombe2003}.
We show how BML matrices are allocated passing the matrix type (variable \verb|"ellpack"| in this case), the element kind (a real kind indicated with the predefined variable \verb|bml_element_real|), and the precision, (in this case a double passed with the variable \verb|dp|). More information about how to use the BML API can be found at \url{https://lanl.github.io/bml/API/index.html}. 
\begin{figure}
\centering
\begin{filecode-0}[]
\begin{lstlisting}[language=Fortran]
!declare BML matrices
type(bml_matrix_t) :: ham,rho,x2
!n:matrix size, m:max. nnz/row
integer :: n, m, dp
real(8) :: thld, tol, nel

dp = kind(1.0d0)

!allocate double precision BML matrices 
call bml_zero_matrix("ellpack", &
&bml_element_real,dp,n,m,ham)
call bml_zero_matrix("ellpack", &
&bml_element_real,dp,n,m,rho)
call bml_zero_matrix("ellpack", &
&bml_element_real,dp,n,m,x2)
...
trx = bml_trace(rho)
do i = 0, niter
  call bml_multiply_x2(rho, x2, thld)
  trx2 = bml_trace(x2)
  if(trx2 .le. nel) then
    ! rho <- 2 * X - X2
    call bml_add(2., rho, -1., x2, thld)
    trx = 2.0 * trx - trx2
  else
    ! rho <- X2
    call bml_copy(x2,rho)
    trx = trx2
  end if
  if (abs(nel-trx) < tol)) exit
end do
...
call bml_deallocate(x2)
...
\end{lstlisting}
\end{filecode-0}
\caption{Fortran example of the SP2 implementation using the BML
ELLPACK-R format with a dropping threshold \texttt{thld}. The algorithm returns the density matrix \texttt{rho} while its parameters 
are the number of electrons \texttt{nel} and the Hamiltonian \texttt{ham}. \texttt{n}, and \texttt{m}
are the total number of orbitals and the maximum number of non-zeros per row.
Only the main operations are shown for brevity.}
\label{sp2}
\end{figure}
Our implementation of various matrix formats are quite mature for CPUs, including
their threaded versions. GPU implementation efforts are ongoing.
Future developments include a distributed version of BML for various matrix types.

A recent effort \cite{adedoyin2020}, focused on optimizing at the multi-threaded level as opposed to modifying the data-structures for performance at the Single Instruction Multiple Data (SIMD) scale. 
Several active and passive directives/pragmas were incorporated that aid to inform the compiler on the nature of the data-structures and algorithms. 
Though these optimizations targeted multi-core architectures, most are also applicable to many-core architectures present on modern heterogeneous platforms. 
\revision{
Herein, we refer to multi-core systems as readily available HPC nodes customarily configured with  approximately 10 to 24 cores per socket at high clock speed  (2.4 to 3.9 GHz) and many-core systems as accelerators or GPUs.
}
Several optimization strategies were introduced including 1.) Strength Reduction 2.) Memory Alignment for large arrays 3.) Non Uniform Memory Access (NUMA) aware allocations to enforce data locality and 4.) the appropriate thread affinity and bindings to enhance the overall multi-threaded performance.

A more in-depth description of the BML library 
and its functionalities can be found in \cite{Bock2018} and the code is available at \url{https://github.com/lanl/bml}.

\subsection{PROGRESS Library}

The
computational cost of solving this eigenvalue problem to compute the DM, scales as O(N$^3$), where N is the
number of atomic orbitals in the system. 
Recursive methods, however, such as SP2 \cite{NiklassonTymczakChallacombe2003}, can compute the density matrix in O(N) operations for a sparse Hamiltonian matrix. 

PROGRESS is a Fortran library that can be used for general purpose quantum chemistry calculations.
It implements several O(N) solvers \cite{Niklasson2016,Negre2016,Mniszewski2019} and is
publicly available at \url{https://github.com/lanl/qmd-progress}. As described above and shown in Figure \ref{bml:stack},
PROGRESS relies entirely on BML for algebraic operations, hence, while electronic structure algorithms and
solvers are outlined in PROGRESS, the mathematical manipulations are all performed in BML.
This
library is currently used by LATTE\revision{, a tight-binding (TB) code specifically developed to perform QMD simulations \cite{LATTE}.
It can also be used with DFTB+, a widely used density functional tight-binding code} \cite{Hourahine2020-hp}.
In TB methods, matrix elements are typically obtained empirically from fits to more accurate calculations or to experiments, rather than being explicitly computed from electronic wave functions.
However, the BML library also can be used for First-Principles codes, in particular for O(N) codes where matrix elements correspond to pairs of localized orbitals \cite{Fattebert2016}.

As was mentioned previously, the appropriate solver to compute the electronic structure depends strongly on the 
type of chemical system. Metals, for example, are difficult to treat since their electronic structure is hard to converge given the delocalized nature of the electrons. The Hamiltonian and DM have different sparsity patterns that will determine the matrix format to use. Hence, there is room for exploring different matrix formats and solvers depending on the type of system. 
The SP2 method, as implemented in the PROGRESS library with the posibility of using BML sparse matrix multiplications, computes the density matrix without diagonalizing the Hamiltonian matrix.  An example SP2 algorithm as implemented in PROGRESS is shown in Figure~\ref{sp2}.
Its computational complexity becomes O(N) for sparse Hamiltonians,
provided a proper thresholding is applied at every iteration. Performance of the PROGRESS library is tested using model Hamiltonian matrices that mimic the actual Hamiltonians for different materials such as semiconductors, soft-matter, and metals. 

\begin{figure}[h]
\includegraphics[width=8cm]{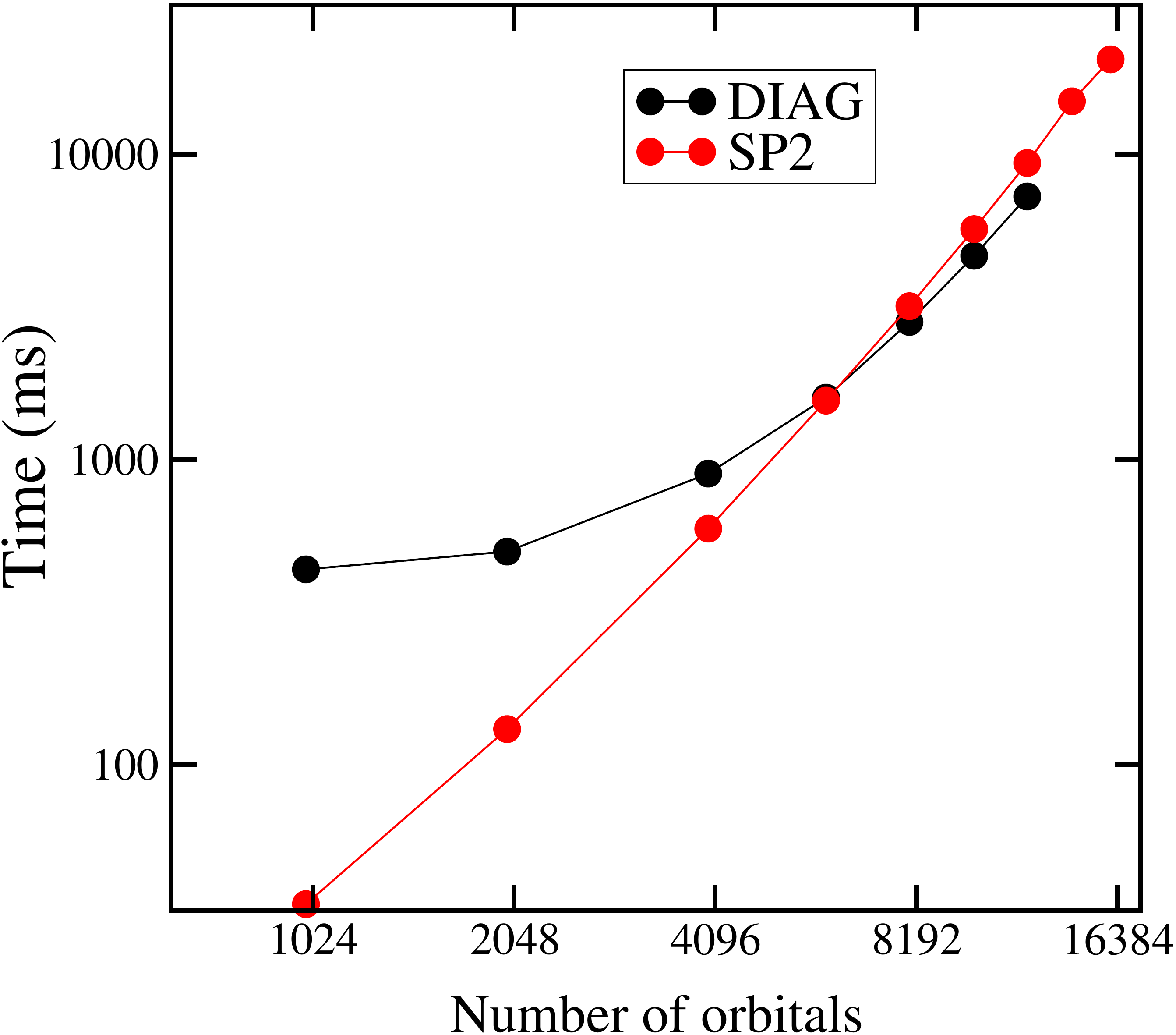}
\caption{GPU performance comparison of two PROGRESS library routines for constructing the density matrix: 
traditional algorithm based on matrix diagonalization (DIAG), and the SP2 algorithm (SP2) based on matrix multiplications. 
Diagonalization is using the NVIDIA cuSOLVER library, while SP2 relies on the MAGMA matrix-matrix multiplication function.
The plot shows the wall-clock time to construct the DM as a function of the number of orbitals. 
Scaling experiments were run on an OLCF Summit node using one V100 NVIDIA GPU. The dense format is used for all BML matrices.}
\label{bml:diag}
\end{figure}
Figure~\ref{bml:diag} shows the performance of two typical PROGRESS routines for constructing the DM on GPU applied to a soft-matter type of Hamiltonian. The standard algorithm for constructing the DM is based on matrix diagonalization (shown in black on the plot of Figure \ref{bml:diag}). The SP2 algorithm (see Figure \ref{sp2}), instead, is based on matrix multiplications (shown in red on the plot of Figure \ref{bml:diag}). 
The computational complexity is O(N$^3$) for both algorithms due to the nature of the matrix operations involved. In both cases these operations scale as O(N$^3$) for dense (unthresholded) matrices. 
Furthermore, in these cases the density matrix is solved exactly since no threshold is used. For systems where the DM becomes dense and where a sparse format cannot be used, the GPU versions of these algorithms are significantly more performant than the corresponding CPU threaded version. 
We also notice that the DIAG algorithm is slower than SP2 for smaller systems (less than 6000 orbitals). This is because the dense diagonalization
algorithm and its implementation on GPUs is not as efficient, in particular for small matrices,
while the SP2 algorithm is dominated by matrix-matrix multiplications which can be implemented very efficiently on GPUs.
For large systems, however, the DIAG algorithm performs slightly better. 

\begin{figure}[h]
\includegraphics[width=9cm]{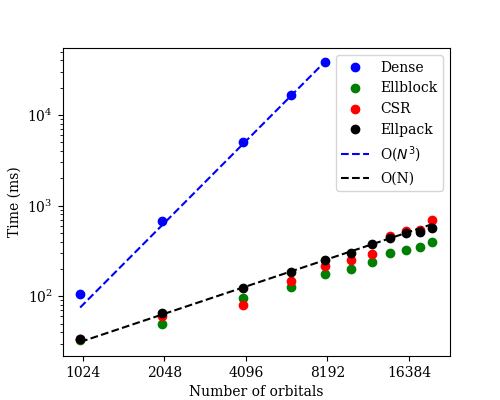}
\caption{Performance of the SP2 algorithm for the construction of the density matrix as implemented in the PROGRESS library. The plot shows the wall-clock time for computing the DM as a function of the number of orbitals. Scaling experiments were run on an OLCF Summit Power 9 node \revision{(using one socket, i.e. 21 cores)} comparing performance using different matrix formats. The threshold was set to $10^{-5}$ for all sparse formats.}
\label{bml:format}
\end{figure}

For systems leading to a sparse DM, O(N) complexity
is achieved by using the SP2 algorithm in combination with a sparse format as is shown in Figure \ref{bml:format}. 
This plot shows the performance of the SP2 algorithm on CPU using different formats (ELLPACK-R, CSR, and ELLBLOCK) applied to a soft-matter type Hamiltonian.  
In this figure we notice a large gain in performance obtained by using sparse formats,
and the O(N) complexity for the range of system sizes analyzed. 
In these cases, the DM is not exact and the error depends on the threshold parameter, regardless of which of the three formats is used, and can be chosen to be sufficient for many practical applications.

Depending on the size of system to be simulated, and the resulting structure of the matrix, optimal time to solution may be obtained from a particular combination of matrix format and hardware choice. The use of PROGRESS and BML allows these choices to be made at runtime, without changing the underlying code structure.

\subsection{PROGRESS/BML Applications}
Our work on PROGRESS/BML is now being used to develop a capability for all-atom QMD simulations of proteins, extending the impact of our ECP work to biomedical research including studies of SARS-Cov-2 proteins. Current classical biomolecular MD simulations typically involve O(10$^4$-10$^5$) atoms and exceed 100 ns in simulation time. Obtaining nanosecond-duration simulations for such systems is currently beyond reach of QMD codes.  The highly scalable NAMD MD code\textemdash a popular choice for biomolecular simulations\textemdash recently incorporated breakthrough capabilities in hybrid quantum mechanical/molecular mechanical (QM/MM) simulations \cite{namdqmmm}, enabling useful simulations with O(10$^1$-10$^2$) of the atoms treated using QM. To extend the size of the QM region to a whole protein with O(10$^3$) atoms, we are now integrating PROGRESS/BML with NAMD, using the LATTE electronic structure code as the QM solver \cite{LATTE}. LATTE uses Density Functional Theory (DFT) in the tight binding approximation which is an established approach for {\em ab initio} studies of biomolecular systems \cite{cui2014}. It combines O(N) computational complexity with extended-Lagrangian Born-Oppenheimer MD, and has achieved a rate of 2.1 ps of simulation time per day of wall clock time for a solvated Trp cage miniprotein system consisting of 8349 atoms \cite{Mniszewski2015}. The combination of LATTE with NAMD therefore is an excellent choice for pursuing nanosecond-duration whole-protein QM/MM simulations.

Our efforts in integrating LATTE with the PROGRESS and BML libraries have significantly benefited the EXAALT ECP project.
Some materials which are the subject of study in EXAALT such as UO$_2$, required a very involved modification of the LATTE code to increase performance. This was made possible by the extensive use of the PROGRESS and BML routines.

\section{PIC Algorithm Development}

The longevity of any software framework is dictated, at least partially, by its ability to adapt to emerging algorithms that may not have existed, nor been foreseen, at the time of the framework's development. In this regard, CoPA has been supporting the development of novel PIC algorithms that can improve and accelerate simulations, with an eye toward their implementation in Cabana.  In addition to exercising Cabana, this also represents a unique opportunity to rethink and develop new algorithms at scale, potentially shortening the often lengthy gap between algorithmic innovation and subsequent scientific discovery.  \revision{These algorithms may connect with the PIC codes in XGC and HACC, described in the next sections.}

The algorithms being developed build mainly on two recent advances in PIC methodology.  The first is the fully implicit PIC algorithm first introduced in \cite{chen2011energy} and subsequently expanded upon in \cite{chen2020semi, chen2015multi, stanier2019fully, chen2012efficient}, among others.  Compared to standard PIC algorithms, these implicit methods enforce exact discrete conservation laws, which are especially important for long-term accuracy of simulations. Their ability to stably step over unimportant but often stiff physical time-scales promises tremendous computational speed-ups in certain contexts.  The second recent advance being leveraged is known as ``sparse PIC'' \cite{ricketson2016sparse, ricketson2018sparse}, in which the sparse grid combination technique (SGCT) \cite{griebel1992combination} is used to reduce particle sampling noise in grid quantities.  This is achieved by projecting particle data onto various different component grids, each of which is well-resolved in at most one coordinate direction.  A clever linear combination of these grid quantities recovers near-optimal resolution, but with reduced noise due to the increased size of the cells - and consequently more particles per cell - in the component grids.  \revision{Quantitatively, use of the SGCT changes the scaling of grid sampling errors from from $O( (N_p \Delta x^d)^{-1/2} )$ to $O( (N_p \Delta x)^{-1/2} | \log \Delta x |^{d-1} )$~\cite{ricketson2016sparse}.  Here, $N_p$ is the total number of particles in the simulation, $\Delta x$ the spatial cell size, and $d$ the spatial dimension of the problem.  For comparable sampling errors, sparse PIC thus reduces the required particle number by a factor of $O(  1/(\Delta x |\log \Delta x|^2)^{d-1} )$.}  

Algorithm development efforts within CoPA have been three-fold.  First, a new asymptotic-preserving time integrator for the particle push \revision{- PIC item 4 in Figure~ \ref{intro:anatomyTS} -} component of implicit PIC schemes has been developed.  This new scheme allows implicit PIC methods to step over the gyroperiod, which often represents a stiff time-scale in strongly magnetized plasmas (e.g. in magnetic confinement fusion devices).  Second, implicit PIC schemes have been generalized to handle a broader class of electromagnetic field solvers \revision{- PIC item 2 in Figure~\ref{intro:anatomyTS}}.  In particular, we show that exact energy conservation can be implemented using a spectral field solve, while previous studies have been mostly focused on finite-difference schemes. Spectral solvers have much higher accuracy given the same degree of freedom than finite-difference schemes, which can have particular advantages in simulating electromagnetic waves (e.g. in laser-plasma interaction applications).  Third, we show that the implicit and sparse PIC methods can be used in tandem, thereby achieving the stability and conservation properties of the former along with the noise-reduction properties of the latter.  \revision{Use of sparse PIC primarily impacts particle depostion and force gather operations - PIC items 1 and 3 in Figure \ref{intro:anatomyTS}.}  Each of these advances is described in turn below.  

In magnetized plasmas, charged particles gyrate around magnetic field lines with frequency $\Omega_c = q B / m$, where $q$ is the magnitude of the particle's charge, $B$ the magnetic field strength, and $m$ the particle mass.  In many scientific applications, the time-scale of this gyration (i.e.\ $\Omega_c^{-1}$) can be orders of magnitude smaller than the physical time-scales of interest. Consequently, it is often too expensive for standard PIC to simulate those applications. Numerous works have circumvented this difficulty by using gyrokinetic models, in which the gyration scale is analytically removed from the governing equations by asymptotic expansion.  However, this approach can become difficult, or breaks down if the approximation is only valid in a portion of the problem domain - scientifically relevant examples include the edge of tokamak reactors, magnetic reconnection \cite{lau1990three}, and field reversed configurations \cite{tuszewski1988field}.  A more effective approach is to derive a time-integrator that recovers the gyrokinetic limit when $\Omega_c \Delta t \gg 1$ while recovering the exact dynamics in the limit $\Delta t \rightarrow 0$.  

Our work \revision{derives just such a scheme.  We present a summary here; the interested reader should refer to \cite{ricketson2020energy} for more detail.  Note that this work} represents an inter-project collaboration with the HBPS SciDAC, \revision{and may be viewed as part of XGC's efforts toward full-orbit capability.}  

\revision{Our new algorithm builds} on prior efforts \cite{brackbill1985simulation, genoni2010fast, parker1991numerical, vu1995accurate} that noticed that standard integrators such as Boris and Crank-Nicolson fail to capture the proper limit when $\Omega_c \Delta t \gg 1$.  In particular, the Boris algorithm \cite{parker1991numerical} captures magnetic gradient drift motion but artificially enlarges the radius of gyration, while Crank-Nicolson captures the gyroradius but misses the magnetic drift.  Some prior efforts cited above derive schemes that capture both, but at the cost of large errors in particle energy.  These energy errors are particularly problematic in the long-time simulations enabled by exascale resources.  

Our new scheme is the first to capture magnetic drift motion, the correct gyroradius, and conserve energy exactly for arbitrary values of $\Omega_c \Delta t$.  The scheme is built on Crank-Nicolson, but adds an additional fictitious force that produces the magnetic drift for larger time-steps, and tends to zero for small time-steps (thus preserving the scheme's convergence to the exact dynamics).  Energy conservation is preserved by ensuring that this fictitious force is necessarily orthogonal to particle velocity, thereby guaranteeing that it can do no work (i.e. mimicking the effects of the Lorentz force).  The scheme shows promising results in various test problems, and implementation in Cabana is expected to help guide the development of effective preconditioning strategies for the necessary implicit solves.  

Our second thrust concerns the field solvers (these are examples of \textit{long-range solvers} - see the section on Cabana above and SWFFT below for additional discussion) in implicit PIC schemes.  In the references above and all other extant implicit PIC work, these solvers are assumed to be based on second-order finite difference approximations of the underlying partial differential equations.  However, there are significant advantages to the use of spectral solvers when treating electromagnetic waves~\cite{vay2013domain,vay2018warp}.  

With these considerations in mind, we have generalized the implicit PIC method to function with spectral solvers without sacrificing the important conservation properties the scheme enjoys.  This is done by adapting the mathematical proofs of energy conservation to accommodate spectral solvers.  The key necessary features are two integration-by-parts identities that must be satisfied by the solver.  A spectral solver \revision{can be made to} satisfy these identities \textit{if} a binomial filter is applied in a pre-processing step.  Such filters are commonly used in PIC schemes to mitigate particle noise, so this requirement is not considered onerous.   

The third prong consists of combining sparse grid PIC schemes with implicit PIC.  As above, the key here is retaining energy conservation.  Because potential energy is computed on the grid and the SGCT introduces a multitude of distinct grids with different resolutions, care is needed even in the definition of a single potential energy quantity.  Having taken this care, we have shown that it is indeed possible to conserve energy exactly in the sparse context.  The resulting method also leverages the advances above by being compatible with spectral field solvers.  

\begin{figure}
    \centering
    \includegraphics[width=0.5\textwidth]{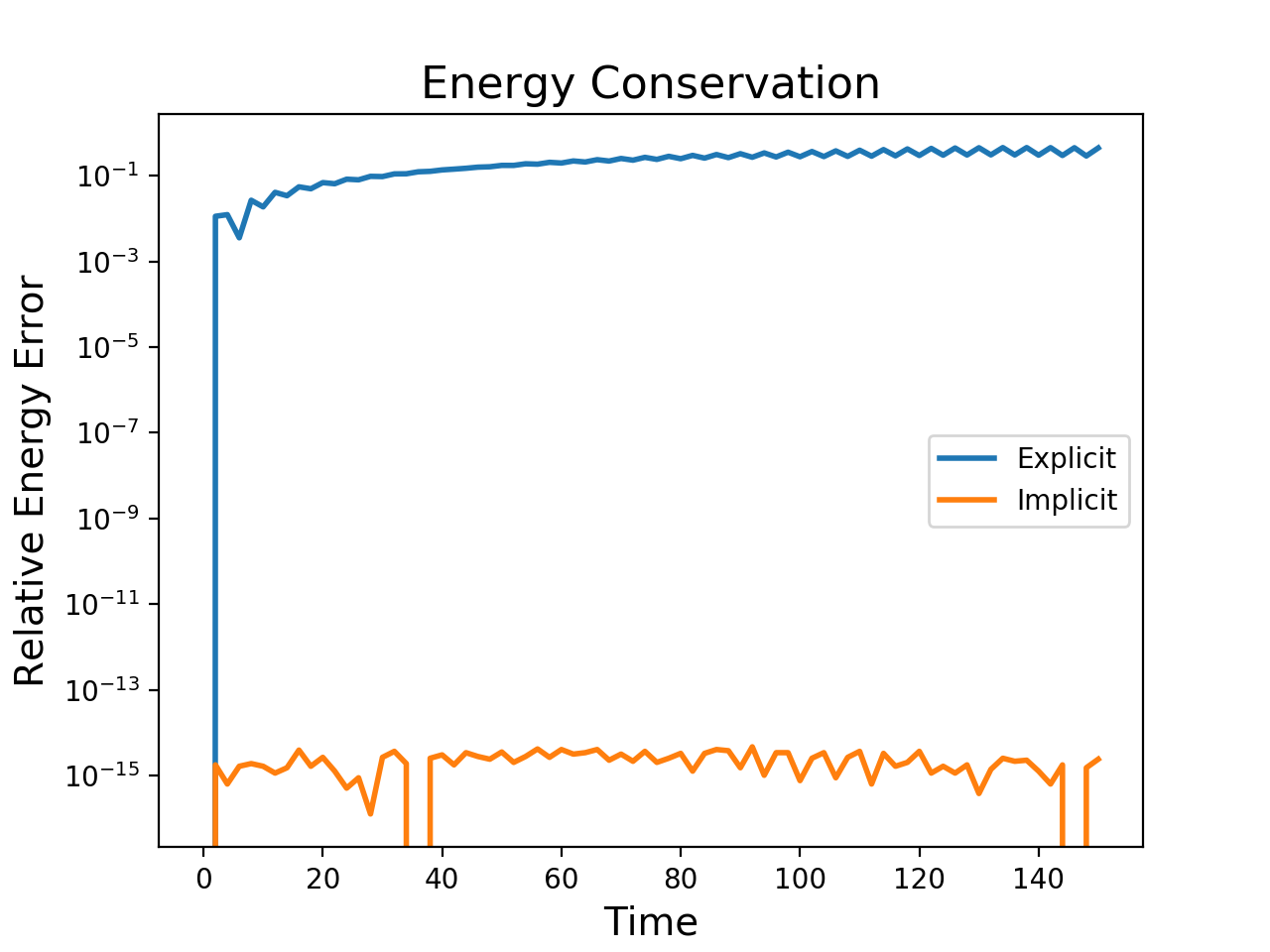}
    \caption{Improved energy conservation properties of implicit sparse PIC (orange) compared to traditional standard PIC \revision{when applied to the diocotraon instability}.  Note that the implicit scheme conserves energy to machine precision.}
    \label{fig:Econs}
\end{figure}

\begin{figure}
    \centering
    \hspace{-1em} 
    \includegraphics[width=0.5\textwidth]{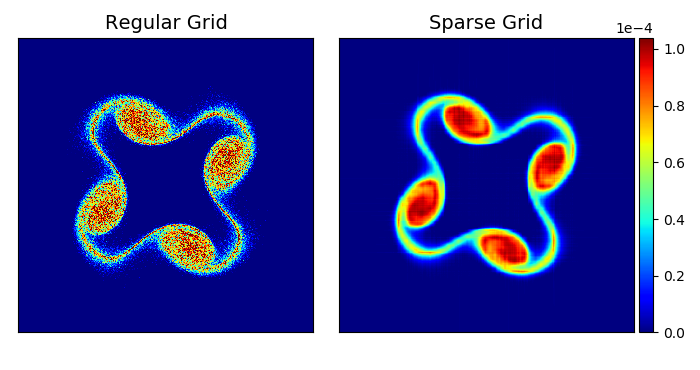}
    \caption{\revision{Comparison of electron number density for the diocotron instability computed by t}wo particle-in-cell schemes using the same number of particles ($2.6 \times 10^6$) and grid resolution ($2048^2$).  On the left, regular grids with explicit time-steps are used.  On the right, the implicit sparse scheme outlined here is used, with immediately visible reduction in particle sampling noise.}
    \label{fig:sparsenoisereduce}
\end{figure}

\revision{Initial tests have} confirmed the theoretically predicted conservation properties, as depicted in Figure \ref{fig:Econs}\revision{, which shows energy conservation up-to numerical round-off for the new implicit scheme applied to the diocotron instability \cite{davidson2001physics}}.  In addition, we illustrate in Figure \ref{fig:sparsenoisereduce} the ability of the sparse grid scheme to dramatically reduce particle sampling noise in the solution.  Future implementation of this method in Cabana poses unique challenges, as its structure is rather different from a typical PIC method.  This is due to the various grids required and the need to perform not only particle-grid and grid-particle interpolations, but also grid-grid interpolations for post-processing.  \revision{As a result, it offers a particularly valuable test-case for the flexibility of the software infrastructure.}  

\section{Application Partners}

\revision{To enable a deep window into how particle applications use the computational motifs, the CoPA co-design center established partnerships with several ECP application development projects. Direct application engagement through deep dives and hackathons has resulted in XGC adoption of Cabana/Kokkos, LAMMPS-SNAP GPU algorithm optimization, and the open source HACC/SWFFT code. Details of these engagements and their impact on exascale readiness are presented below. Kernels can easily be extracted and explored through proxy apps leading to performance improvements. CoPA's inter-dependencies with ECP \revision{ST} library projects\revision{, which provide common software capabilities,} has also led to improvements and additions. Details of these engagements are described below.}

\section{XGC and WDMApp}


\revision{In this section, we show how the Cabana library has been utilized to enable the fusion WDMApp PIC code XGC to be portable while preserving scalibility and performance. In the Anatomy of a Time Step Figure~\ref{intro:anatomyTS}, XGC fits into the PIC sub-motif.  The particle movement in the particle resorting step has been minimized to avoid MPI communications.  Instead, the particle remapping step is heavily utilized in each particle cell independently, which is an embarrassingly parallel operation.} 

The ECP Whole Device Model Application (WDMApp) 
project's aim is to develop a high-fidelity model of magnetically confined fusion plasma that can enable better understanding and prediction of ITER and other future fusion devices, validated on present tokamak (and stellarator) experiments. In particular, it aims for a demonstration and assessment of core-edge coupled gyrokinetic physics on sufficiently resolved time-scales to study formation of the pedestal, a physical phenomenon essential to ITER's success but whose mechanisms are still not well-understood. The WDMapp project involves coupling a less expensive code (GENE continuum code or GEM PIC code, solves for perturbed parts only), which models the tokamak's core, with a more expensive code, XGC (obtains total 5D solution), which is capable of modeling the edge of the device plasma where the computational demands are highest. Performance of the coupled WDMApp code is expected to be dominantly determined by XGC. Performance optimization of XGC is essential to meet the exascale demands.

XGC is a Fortran particle-in-cell code used to simulate plasma turbulence in magnetically confined fusion devices \cite{Sku18}. It is gyrokinetic, a common plasma modeling approach in which velocity is reduced to two dimensions (parallel and perpendicular to the magnetic field), thus reducing total model complexity from 6D to 5D. Markers containing information about the ion and electron particle distribution functions are distributed in this phase space. In a given time step, particle position is used to map charge density onto an unstructured grid. The charge density is solved to determine the global electric field, which in turn is used to update (``push'') particle position for the next time step.  Particle velocity is also mapped onto an unstructured grid to evaluate the velocity space Coulomb scattering in accordance with the Fokker-Planck operator.

Electron position must be updated with a much smaller time step than ions due to their high relative velocity. They are typically pushed 60 times for every ion step (and field solve), and as a result the electron push is by far the most expensive kernel in XGC.

In the past, several versions of the electron push kernel were developed that maximized performance on specific hardware. XGC maintained a CUDA Fortran version of the electron push kernel optimized for the previous Oak Ridge supercomputer Titan; an OpenMP version that vectorizes and performs well on CPUs; and an unvectorized OpenMP version for use as a cleaner reference.

In addition to the basic time step cycle described above, XGC also has source terms including a Fokker Planck collision solver on each grid node as briefly mentioned above, which is the second most computationally expensive kernel after the electron push. This kernel offloads work to GPU with OpenACC if available, or uses OpenMP if on CPU. Utilizing multiple offloading programming models in the same simulation poses additional challenges when adapting the code to new platforms and compilers. For example, on Summit only one available compiler (PGI) supported both OpenACC and CUDA Fortran.

To prepare for exascale architectures, XGC is in the process of significant restructuring. Instead of multiple code bases and offloading programming models, it is being rewritten to use Kokkos and Cabana and to strive toward a single maintainable, flexible codebase that performs well on all relevant architectures \cite{Scheinberg19}.

\subsection{Kokkos/Cabana Implementation}

Since XGC is written in Fortran, utilizing Kokkos and Cabana posed the additional challenge of Fortran-C++ interfacing. Our initial goal was to use these libraries without significant changes to the main code or to the individual kernels to be offloaded. We developed an initial such implementation, in which the XGC main code would call a C++ subroutine that wrapped a Kokkos parallel\_for that launched a kernel that looped over particles and called the necessary Fortran kernel. Kokkos was therefore restricted to a thin interface that managed kernel launching. The Fortran kernel itself had to be modified with preprocessor macros which directed the compiler to compile the code for CPU or GPU as specified; under the hood, CUDA Fortran was still used for GPU offloading.

There were several downsides to this approach. First, it restricted the Kokkos and Cabana features available for use, instead often necessitating custom features for memory management and host-device communication. Second, reliance on Fortran modules often made proper encapsulation difficult. Third, it was unclear if the approach could be easily extended to platforms with AMD or Intel GPUs where no foolproof equivalence to CUDA Fortran would be available. For these reasons, we instead opted to convert XGC into C++, beginning with the major kernels that require offloading, and gradually converting the remaining code. With this new approach, we are better able to utilize the strengths of Kokkos and Cabana by relying on them for memory management, host-device communication, etc.

Due to the piecemeal approach to the code conversion, many data structures on the CPU are still allocated on the Fortran side.  At each time step, the particles are rearranged into an array of structures and sent to the GPU as a Cabana AoSoA object. Other data residing in Fortran arrays are wrapped in unmanaged Kokkos Views and can then be copied to Views on the GPU. This method was found to be the least disruptive means of interfacing as the code is gradually converted to C++.

Within kernels that loop over particles, an inner loop is also present, with a range of 1 on GPU and a pre-compiled length (32 by default) on the CPU. These inner loops are mostly vectorized if compiled on CPU, and loop over particles within a single structure of arrays from the AoSoA while the outer loop (the parallel\_for) loops over structures with OpenMP. The result is a single code base that vectorizes if compiled for CPU and coalesces if on GPU.

\subsection{Results}

The most expensive operation, the electron push, is now in C++ and offloaded using Kokkos, as well as electron charge deposition and sorting. Since the ion push is independent from the electron push and is still CPU-only, it is performed asynchronously while the electron push is performed on GPU.

\begin{figure*}[h]
\includegraphics[width=16cm]{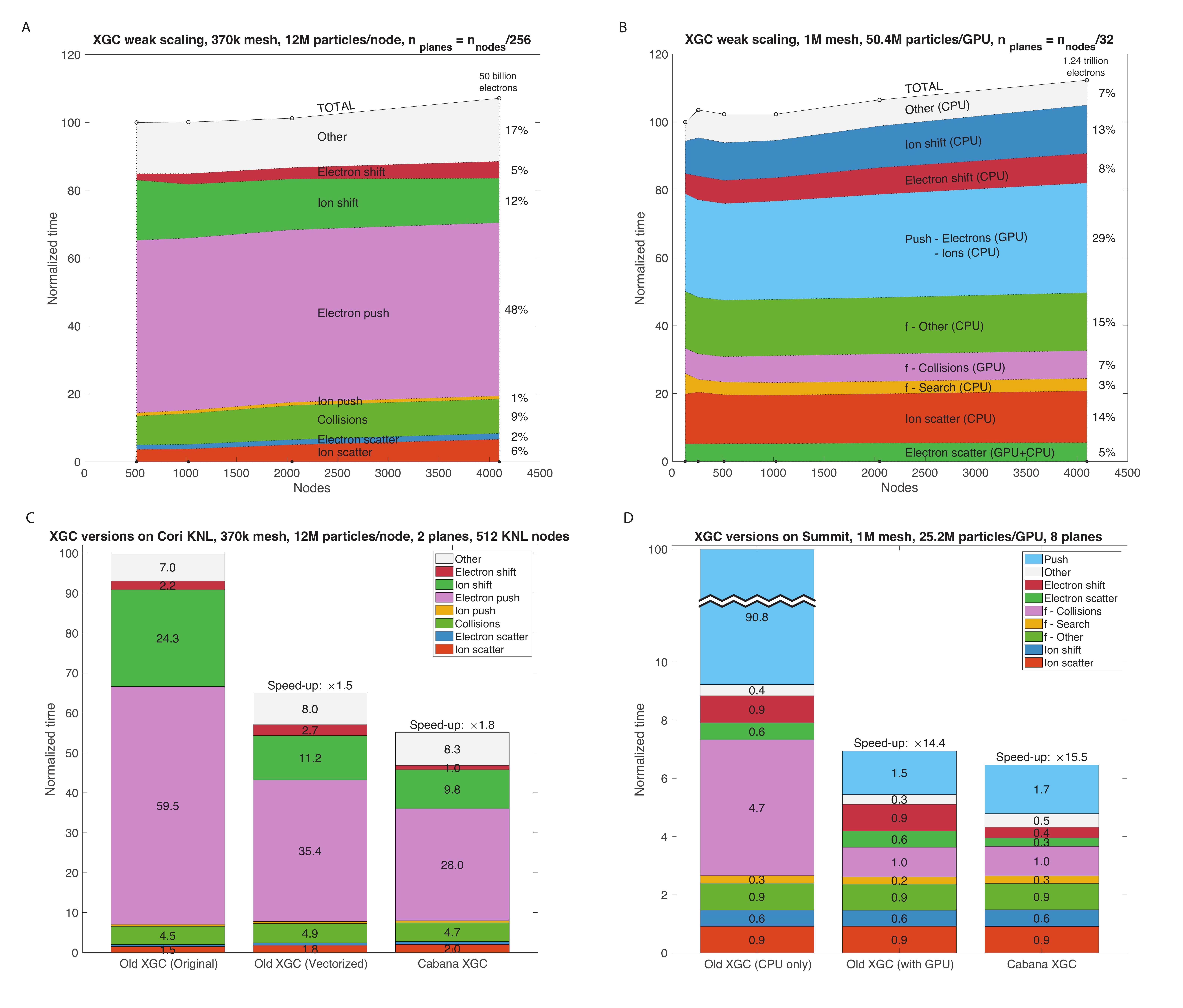}
\caption{\revision{Performance of the whole production XGC code} evaluated \revision{on the KNL partition of Cori (A, C), and on Summit (B, D). (A, B)}: Weak scaling studies on both machines demonstrate that supercomputer-scale simulations can be done with the Cabana version without loss of performance. (C, D): The Cabana version is compared against previous versions of XGC: an unvectorized OpenMP version and a vectorized OpenMP version on Cori \revision{(C)}, and the OpenMP version and a CUDA Fortran version on Summit \revision{(D)}. \revision{We caution here that the colors and legends are different between the KNL partition of Cori (A, C) and Summit (B, D).} In both cases, the Cabana implementation of the expensive electron push kernel performs about as well as the previous, architecture-specific implementations. \revision{Ion-push color in (C) is not visible because the wall-time of the kernel is negligibly short in the KNL partition of Cori.}}
\label{XGC_plot}
\end{figure*}

A scaling study and comparisons between the different code bases were conducted on both Summit and Cori (KNL) supercomputers. These tests used simulation parameters and size comparable to those used in scientific production. The new code was found to weak-scale well on both machines (Figure~\ref{XGC_plot}A-B). Performance on Cori was found to be similar to the performance of the vectorized Fortran code, while performance on Summit is also similar to the CUDA Fortran version of the code (Figure~\ref{XGC_plot}C-D). In fact, the Kokkos/Cabana version outperformed previous versions; however, the improvements cannot be entirely attributed to this, since minor algorithmic and structural changes occurred during the porting process.

We conclude that adopting Kokkos and Cabana enabled us to consolidate to a single codebase that is portable to diverse architectures without sacrificing performance.

\subsection{Exascale Outlook}

Conversion of the remaining XGC kernels into C++ is underway. In addition to offloading more XGC kernels, experimentation with more Cabana features (sorting, inter-GPU particle exchange, etc.) will be performed. This may prove useful particularly as more data will be resident on GPU on exascale architectures.

The collision kernel has been converted and offloaded with Kokkos, though performance results are not yet available. With the new collision kernel, OpenACC will no longer be needed and XGC will rely solely on Kokkos for GPU offloading.

\section{ExaSky}


The ExaSky ECP project 
focuses on extreme-scale cosmological simulations targeted at next-generation sky surveys that observe across multiple wavebands. The simulations follow the development and evolution of cosmic structure in an expanding universe, including not only the effects of gravity, but also gas dynamics and a number of astrophysically relevant processes such as radiative cooling, star formation, and various feedback mechanisms, several of which are treated via phenomenological subgrid models.

Cosmological simulations have a vast dynamic range in space, approximately six orders in magnitude, and the corresponding demands on time and density resolution are very severe. ExaSky uses two codes, HACC (Hardware/Hybrid Accelerated Cosmology Code)~\cite{HACC} and Nyx~\cite{Nyx}; HACC uses tracer particles for both dark and ordinary matter (`baryons'), whereas Nyx uses an Eulerian adaptive mesh refinement (AMR) based method for the gas dynamics. Nyx is strongly coupled to methods being developed by the AMReX ECP co-design center,
whereas HACC, because it is essentially a Lagrangian, particle-based code framework, has strong ties to CoPA.
\revision{In Figure~\ref{intro:anatomyTS}, HACC represents a combination of sub-motifs, where PIC methods are used for a Poisson solver to calculate gravitational forces over large distances, and MD-like methods on nearby particles are used to evaluate local contributions to the gravitational force. More details about HACC's gravitational force-splitting are given in the next section.}

\subsection{HACC}
HACC solves the the 6-D Vlasov-Poisson equation in an expanding universe~\cite{peebles} and includes gas dynamics via a new SPH scheme, CRK-SPH (Conservative Reproducing Kernel SPH), an effectively higher-order method that overcomes many of SPH's known problems, while maintaining its advantages~\cite{frontiere}. HACC's gravity solver splits the gravitational force computation into two parts, a long-range Poisson solver based on a high-order hybrid spectral method, and matched short-range solvers that are designed to be separately optimized for different architectures (direct particle-particle, tree, fast multipole). HACC's long-range solver is essentially a PIC method that actively leverages the use of a large, distributed FFT to minimize indirection, reduce particle noise, isotropize the force kernel, and compactly implement higher-order methods for particle deposition and force computation. Time-stepping is performed via an adaptive split-operator, symplectic method that uses subcycling for increased temporal resolution for the dynamics associated with the short-range force.  HACC's Poisson solver is unusual in that it uses error compensation in the Fourier domain to effectively increase the order of the solver even though the particle-grid interaction is only kept to first nontrivial order (i.e., CIC deposition and interpolation). Details are given in~\cite{HACC}.

HACC has its own dedicated, distributed 3D FFT, SWFFT (see below), which has been made publicly available under CoPA. The short-range gravity and hydro solvers comprise the most computationally intensive kernels within HACC and are heavily performance-optimized on a number of architectures. These kernels are highly compact and are excellent candidates to test and exploit the performance portability possibilities using the Cabana framework. As a \revision{deliberate} result of HACC's design, 95\% of the code does not change as one runs on different platforms (e.g., CPU or CPU+GPU systems), a feature which greatly aids in implementing different performance-portable solutions. \revision{Because of the isolation of the computational work into a finite number of compact kernels, a very high level of targeted performance optimization is possible, which would not be the case with the use of generic external libraries. Additionally, the algorithms used are also tied to the architecture as an instance of ``software co-design'' so the dependencies are not static. Finally, as HACC is often used as a benchmark code on emerging architectures, performant libraries often do not exist on these platforms.}

Future work envisaged for HACC is a proxy app based on Cabana and a general long-range solver implemented in Cabana that uses high-order spectral gradients. In addition, as a test of performance portability, we envisage building a short-range gravity kernel in Cabana that can interface with the rest of the HACC code. In this case we can run the full code with a compact, localized modification.

\subsection{CosmoTools}
CosmoTools is the analysis framework associated with HACC. \textit{In situ}, co-scheduled, and offline analyses associated with HACC are complex and computationally demanding in their own right, and are as important as running the underlying simulations. Because the analysis methods are diverse, performance portability, and especially the ability to use accelerators are both key issues for CosmoTools.

In the ECP context, \textit{in situ} analysis is of particular importance. Some algorithms in CosmoTools can be built on primitives used by the solver, whereas others, such as neighbor-finding and other clustering-based measurements are unique to CosmoTools; implementation of the latter class of methods often requires the use of efficient graph algorithms. Work is ongoing with the ArborX team~\cite{arborx} to implement new algorithms for clustering analyses (e.g., DBSCAN, N-point correlation functions) on GPUs with promising initial results having been obtained.

\subsection{SWFFT}

\begin{figure}[h]
\includegraphics[width=8cm]{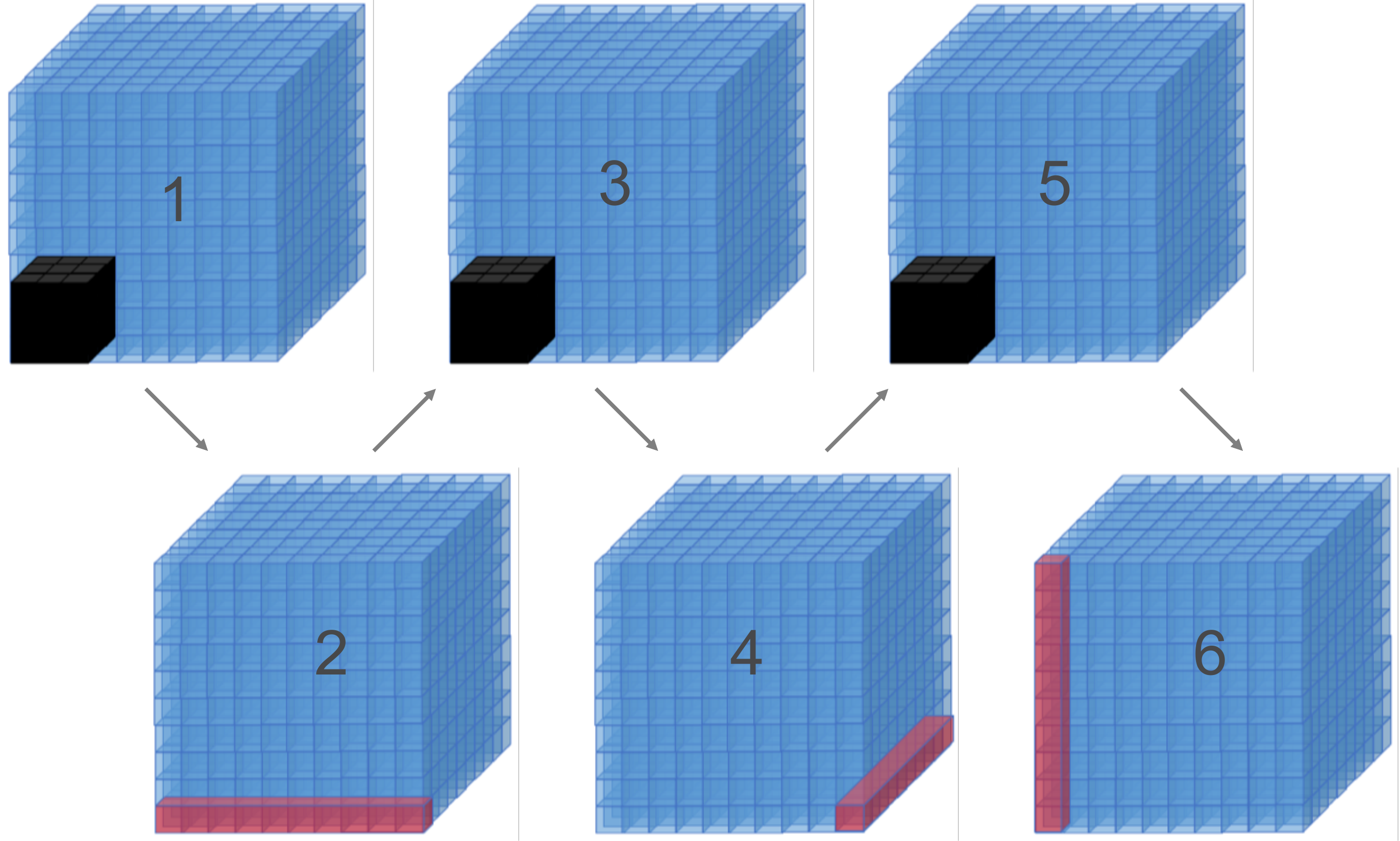}
\caption{SWFFT decompositions and communication pattern.}
\label{swfft-decomposition}
\end{figure}

HACC's performance and scaling requirements involve running very large 3D FFTs ($n_g^3$ grids, where $n_g \gtrsim 10^4$) distributed across a potentially very large number of MPI ranks ($n_R \gtrsim 10^6$) in order reach the desired dynamic range of the long-range gravitational force via a Poisson solver. A common first approach to a distributed-memory 3D FFT is to divide the grid among ranks along one dimension at a time, creating a 1D ``slab'' decomposition, but this approach can only work if the number of ranks does not exceed the number of grid vertexes along one dimension ($n_R \leq n_g$). In order to scale to more ranks, HACC's 3D FFT employs a 2D ``pencil'' decomposition, where ranks are distributed across one face of the grid at a time, relaxing the constraint on the number of ranks relative to grid size ($n_R \leq n_g^2$). HACC's particle operations are sensitive to the ratio between surface area and volume on each rank, so the deposition of particle information onto a grid occurs in a 3D ``brick'' decomposition. HACC's 3D FFT requires grid data to be redistributed between the 3D brick decomposition and each of three 2D pencil decompositions (${\rm 2D}_x, {\rm 2D}_y, {\rm 2D}_z$) where the actual 1D FFTs are performed one dimension at a time. HACC's 3D FFT has implemented this process by going back to the 3D brick decomposition in between all of the pencil decompositions (see Figure~\ref{swfft-decomposition}), so the only communication routines are those that go back and forth between 3D and 2D decompositions. The HACC development team maintains an open source version of this 3D FFT as the SouthWest Fast Fourier Transform (SWFFT, \url{https://xgitlab.cels.anl.gov/hacc/SWFFT}). SWFFT is implemented as an out-of-place transform on double-precision complex grid data. The low level communication is implemented in C, the native high-level FFT interface is implemented as header-only C++, and a Fortran FFT interface is also supported.
\revision{Currently there are a few minor differences in the API between SWFFT and HACC's internal 3D FFT, but the codes are functionally the same.}

\begin{figure}[h]
\includegraphics[width=8cm]{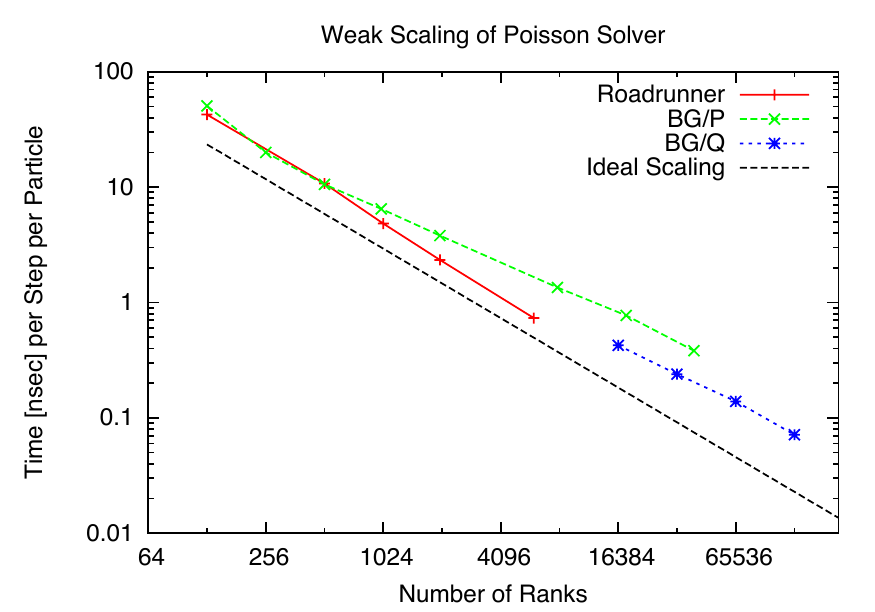}
\caption{SWFFT weak scaling, reproduced from the SC12 publication~\cite{gb12}.}
\label{swfft-scaling}
\end{figure}

SWFFT's implementation and performance characteristics are driven by HACC's requirements, and the primary goal is excellent weak scaling in memory-limited regimes. An advantage of SWFFT's communication pattern is that the number of rank pairs that must exchange data scales as the cube-root of the total number of ranks ($n_R^{1/3}$). For a communication pattern where data is exchanged directly between pencil decompositions, the number of rank pairs that must exchange data scales as the square-root of the total number of ranks ($n_R^{1/2}$). HACC can maintain a relatively small number of large messages as the number of MPI ranks becomes large, though there are several more communication stages than a direct pencil-pencil communication pattern, so this can emphasize robust weak scaling over absolute minimum latency. Figure~\ref{swfft-scaling} shows the scaling of HACC's Poisson solver, where each Poisson solve involves four 3D FFTs - one forward, three backward for force components using spectral gradients. The largest HACC simulation so far used a $15230^3$ grid on 1,572,846 MPI ranks on LLNL's Sequoia IBM Blue Gene/Q system, and each FFT took ${\sim}10$ seconds to complete. In addition to the source and destination grid memory, SWFFT uses send and receive buffers to reorganize data into messages, but the fractional overhead of those buffers scales as $n_R^{-1/3}$ and becomes smaller at larger scales.

The stand-alone SWFFT code was developed to serve as a MiniApp that represented the dominant communication workload in HACC, and also as a potential tool for use in solvers in other applications. Through CoPA and ExaSky, an experimental version of Nyx implemented a gravitational solver based on SWFFT. For Nyx, a branch of SWFFT was created with additional flexibility in mapping 3D sub-volumes to MPI ranks, and that branch will be re-integrated into the main branch and used to support a new memory-balancing mode in HACC. We are also exploring integrating SWFFT as a backend FFT for solvers written in CoPA's Cabana framework. SWFFT has already demonstrated scaling up to ${\sim}15,000^3$ grids on ${\sim}1.5M$ MPI ranks, and on exascale systems HACC plans to use $20,000^3 - 30,000^3$ grids. By maintaining the stand-alone open source SWFFT and participating in ECP, we hope to help other applications and science domains that could benefit from using extremely large FFTs on exascale systems.

\section{Improving GPU performance of a machine learned potential for MD}
The EXAALT project within ECP seeks to extend the accuracy, length, and time scales of materials science simulations to model plasma-facing metals used in future fusion reactors like ITER.  One method to extend time scales is to run up to millions of small molecular dynamics (MD) simulations (1K to 1M atoms each) and use the parallel replica dynamics (PRD) algorithm as encoded in the ParSplice program \cite{ParSplice} to stitch them together into statistically accurate long timescale trajectories.  To accurately model defects in metals surfaces bombarded with plasma ions, each replica uses the SNAP machine-learned (ML) interatomic potential \cite{Thompson2015}, available in the LAMMPS MD code \cite{Plimpton1995} (\url{https://lammps.sandia.gov}). The ability to run the full-scale model on an exascale machine for long timescales thus depends on the performance of SNAP on one or a few GPUs when simulating a small system (one replica in the PRD ensemble).

A Kokkos version of the SNAP potential was originally implemented in the ExaMiniMD proxy app (\url{https://github.com/ECP-copa/ExaMiniMD}) and then ported to LAMMPS. At the time ECP began, the fraction-of-peak performance for SNAP for this baseline version was very low on GPUs. To address this concern, a collaboration between EXAALT, CoPA, NERSC/NESAP, Cray, and NVIDIA was formed. A new proxy app version of the SNAP model, called TestSNAP, was created (\url{https://github.com/FitSNAP/TestSNAP}).

TestSNAP is a serial code derived from the parallel CPU version of SNAP in LAMMPS.  It is a good proxy in terms of memory and computational costs. \revision{It computes step 3 of the MD sub-motif in Figure~\ref{intro:anatomyTS}, which dominates all other parts of the timestep for a simulation using SNAP.}  Importantly, the isolation of the SNAP algorithm in the proxy code made it possible to rapidly experiment with different formulations of the high-level algorithm as well as low-level optimizations such as data structure alterations or loop reordering.  The proxy also includes a correctness check which was very helpful to insure changes did not alter the numerical results.  The team used the proxy to explore a variety of GPU strategies, first using the OpenACC and CUDA programming models, and then Kokkos.  Improvements made in TestSNAP were ported back to the Kokkos version of SNAP in the production LAMMPS code.  Further improvements were also implemented directly in LAMMPS \cite{snap_perf}.

The following optimizations improved both CPU and GPU performance  of the SNAP potential in LAMMPS:

\begin{enumerate}

\item Altered the structure of the SNAP equations
  to avoid duplicate computations
  in different terms as well as the order of summations by using an adjoint refactorization.  This enabled a
  dramatic reduction in the flop count, as well as reduced memory footprint and memory access count.

\item Flattened jagged multi-dimensional arrays which further reduced memory use.

\item Symmetrized data layouts of certain matrices, which reduced memory overhead and use of thread atomics on GPUs.

\end{enumerate}

These optimizations were GPU specific:

\begin{enumerate}
\setcounter{enumi}{3}

\item Broke up one large kernel into multiple kernels.  This
  reduced register pressure, but also greatly increased memory use as intermediate quantities needed to be stored between kernel launches. However, with other optimizations, the net effect was a large reduction in memory use with reduced register pressure.

\item Reversed the order of per-atom and per-neighbor loops.

\item Optimized the memory data layout for the chosen access patterns (e.g. column-major vs row-major).

\item Changed the memory data layout of an array between kernels via transpose operations.

\item Refactored loop indices and data structures to use complex numbers and multi-dimensional arrays instead of arrays of structs.

\item Refactored some of the kernels to avoid thread atomics and use of global memory.

\item Judiciously used Kokkos hierarchical parallelism and GPU shared memory.
 
 \item Fused a few selected kernels, which helped eliminate intermediate data structures and reduced memory use.

\item Added a new memory data layout inspired by Cabana, which enforced perfect coalescing and load balancing in one of the kernels.

\item Pre-computation of certain parameters.

\end{enumerate}

\begin{figure}[h]
\includegraphics[width=8.5cm]{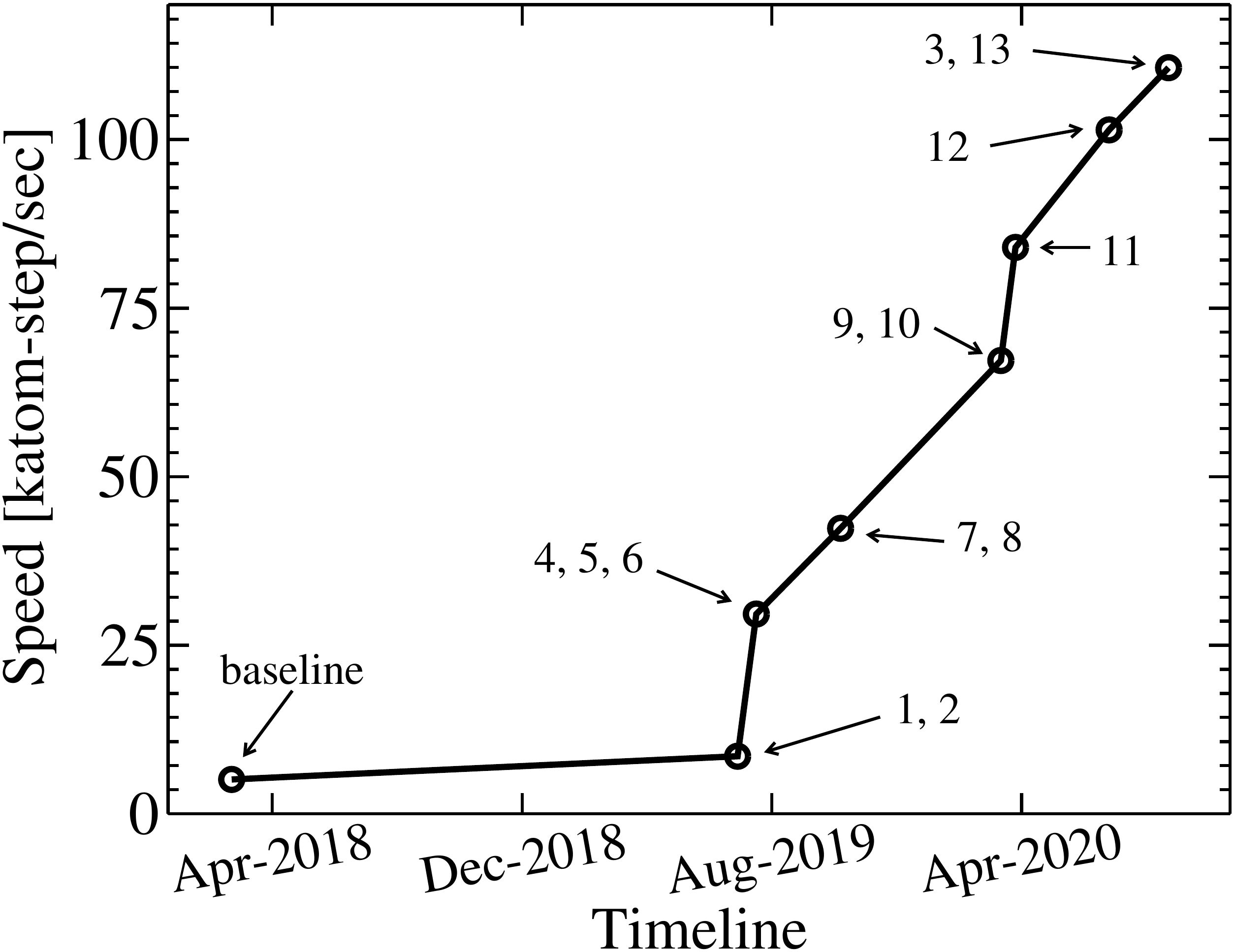}
\caption{22x improvement over time of SNAP performance in LAMMPS on an NVIDIA V100 GPU. The numbered data points refer to specific optimizations in the CPU/GPU and GPU lists.}
\label{snap}
\end{figure}

Figure \ref{snap} shows the effect of these optimizations on the SNAP potential performance over time for the EXAALT benchmark problem running on a single NVIDIA V100 GPU on OLCF Summit. For the original Kokkos version of SNAP in LAMMPS, the performance was 5.09 Katom-steps/s per GPU (Katom-steps = 1000s of atom-steps).  With the improvements listed above, the new performance is now 110.7 Katom-steps/s per GPU, which is a $\sim$22$\times$ speedup.

The algorithmic improvements have also been implemented in the CPU (non-Kokkos) version of SNAP in LAMMPS, with the exception of the third item in the CPU/GPU list. Running on 36 MPI ranks of a dual-socket Intel Broadwell CPU, these changes increased the performance of the CPU version of SNAP by a factor of $\sim$3 for the same benchmark.

\section{Summary}


Library efforts, algorithm development, and interactions with particle applications represented within CoPA all contribute to our co-design process and strategy. 
\revision{The anatomy of a timestep for particle applications (Figure~\ref{intro:anatomyTS}) provides a window into the scope of the CoPA Co-design Center.
The computational kernels requiring optimization for exascale computing are associated with the nature of particle interactions.
Applications with short-ranged, long-ranged, and particle-grid interactions are addressed within the Cabana library.
While applications requiring a quantum mechanical description of interactions are addressed within the PROGRESS/BML libraries.
Inclusion of expertise and application partners representing all the sub-motifs has allowed us to understand and create these libraries as well as proxy apps of interest for short-range MD, long-range MD, PIC, and QMD applications.} 
Success is measured 
by the use of these products within both ECP and non-ECP projects.
We close by highlighting some lessons learned, followed by impacts within ECP and the broader community.

Important lessons learned include:
\begin{enumerate}
\item Many times over we have discovered the benefits of proxy apps for rapid prototyping of different ideas and speedup of the performance optimization process.
\item Improving performance on GPUs requires multiple approaches including optimizing data layout, coalescing memory accesses, increasing arithmetic intensity, and using profiling to guide optimizations. Gains can come from both improving the algorithm as well as improving the implementation. 
\item Co-design teams of domain scientists, computational scientists, and expert programmers in hardware-specific languages and programming models, working together, proved beneficial to design and optimization efforts. 
\item Focused hackathon sessions proved highly productive for small teams over short timeframes, collaborating on algorithms, implementations, and benchmarking.
\end{enumerate}

Impacts as successes with our application partners \revision{across all sub-motifs} include:
\begin{enumerate}
\item WDMApp/XGC is transitioning from Fortran to C++ using Kokkos/Cabana, replacing much of their code with Cabana kernels. The result will be a single flexible codebase with performance portability across relevant architectures.
\item EXAALT/LAMMPS, as part of a co-design team, was able to improve the performance of their SNAP ML model by $\sim$22$\times$.
\item Integration of the PROGRESS/BML QMD capability, the LATTE electronic structure code, and the NAMD MD code, has enabled hybrid QM/MM simulations of proteins. This capability will extend the impact of our ECP work to biomedical research including studies of SARS-Cov-2 proteins.
\end{enumerate}
Library efforts have influenced improvements in a number of the ECP ST libraries, such as Kokkos, heFFTe, ArborX, and others. 
CoPA's library co-design capability allows for integration into existing \revision{particle} applications, as well as creation of new applications as we continue on the road to exascale.



\begin{acks}
This work was performed as part of the Co-design Center for Particle Applications, 
supported by the Exascale Computing Project (17-SC-20-SC), a collaborative effort
of the U.S. DOE Office of Science and the NNSA. 
Assigned: Los Alamos Unclassified Report (LA-UR) 20-26599.

This work was performed at Argonne National Laboratory under the U.S. Department of Energy contract DE-AC02-06CH11357,
Lawrence Livermore National Laboratory under U.S. Government Contract DE-AC52-07NA27344,
Oak Ridge National Laboratory under U.S. Government Contract DE-AC05-00OR22725, Princeton Plasma Physics Laboratory under U.S. Department of Energy contract DE-AC02-06CH11357 with Princeton University, Los Alamos National Laboratory, and at Sandia National Laboratories.

Los Alamos National Laboratory is operated by Triad National Security, LLC, for the National Nuclear Security Administration of the U.S. Department of Energy (Contract No. 89233218NCA000001).


Sandia National Laboratories is a multimission laboratory managed and operated by National Technology and Engineering Solutions of Sandia, LLC., a wholly owned subsidiary of Honeywell International, Inc., for the U.S. Department of Energy's National Nuclear Security Administration under contract number  DE-NA-0003525.

This research used resources of the Oak Ridge Leadership Computing Facility (OLCF), the Argonne Leadership Computing Facility (ALCF), and the National Energy
Research Scientific Computing Center (NERSC), supported by DOE under the contract numbers DE-AC05-00OR22725, DE-AC02–06CH11357, and DEAC02-05CH11231, respectively.


This paper describes objective technical results and analysis. Any subjective views or opinions that might be expressed in the paper do not necessarily represent the views of the U.S. Department of Energy or the United States Government.
\end{acks}

\bibliographystyle{SageH}
\bibliography{CoPA_IJHPCA}


\begin{biogs}
Susan M. Mniszewski is a Senior Scientist at Los Alamos National Laboratory and PI for the Co-design Center for Particle Applications (CoPA). She is also a Co-PI for a Quantum Computing LDRD Project. 
She has contributed to the High Performance Computing PROGRESS/BML libraries for quantum molecular dynamics (QMD). Her research interests include new algorithm approaches for material science applications, machine learning, and novel computing (quantum, neuromorphic).

James Belak is a Senior Scientist at Lawrence Livermore National Laboratory and Co-PI for the CoPA co-design center. His career has centered around the application of High Performance Computing to equilibrium and non-equilibrium problems in Materials Physics.

Jean-Luc Fattebert is a Staff Scientist at Oak Ridge National Laboratory, working on the BML and PROGRESS libraries. His research interest is on computational algorithms to solve problems in materials sciences, chemistry and biology, from the atomistic scale to the mesoscale. 

Christian F. A. Negre is a Scientist at Los Alamos National Laboratory. He has spent most of his career working as a computational chemist studying optical properties of metallic nanoparticles, absorption spectra of organic molecules, interfacial electron and energy transfer between molecules and semiconductors, and molecular electronics. Dr. Negre is now developing techniques to improve the performance of quantum-based molecular dynamics simulations (QMD) focusing on methods to solve problems in the field of applied theoretical chemistry. 

Stuart Slattery is a Computational Scientist at Oak Ridge National Laboratory where he is Team Lead for Scalable Algorithms and Applications. His work focuses on scalable algorithms and performance portable software for applications in advanced manufacturing and nuclear engineering. 

Adetokunbo A. Adedoyin is a Scientist at Los Alamos National Laboratory specializing in scientific application performance on state-of-art and  future computer architectures. Prior to LANL, he served as a Computational Physicist at the University of Notre Dame specializing in constitutive modeling of advanced reactive materials at the macro- and meso-scopic scale.

Robert Bird is a Scientist at Los Alamos National Laboratory, who specializes in the development of performance-portable code and algorithms for next generation compute platforms. His work focuses primarily on particle methods, but also extends to other areas. Within CoPA, he is both a core Cabana developer and a plasma-PIC specialist.

CS Chang is a Managing Principal Physicist at Princeton Plasma Physics Laboratory. He is the head of the SciDAC Partnership Center for High-fidelity Boundary Plasma Simulation and the Co-Lead for Science of the ECP WDMApp project.  He is also the leader of the international XGC particle-in-cell code development team.  His interest is focused around the extreme-scale HPC study of the non-local, nonlinear, multiscale plasma turbulence and transport.

Guangye Chen is a Scientist at Los Alamos National Laboratory. His research interests include computational plasma physics, novel algorithm development, scientific high-performance computing, and software development.

St\'ephane Ethier is a Principal Computational Scientist at the Princeton Plasma Physics Laboratory (PPPL) and co-head of the Advanced Computing Group. His work focuses on high performance computing on large-scale systems, particle-in-cell methods for magnetic fusion research, GPU programming, data management, and other related fields. He is a member of the ECP Whole Device Modeling Application project, as well as CoPA.

Shane Fogerty is a Scientist at Los Alamos National Laboratory. His research spans topics related to performance-portable computational methods for scientific simulation software. He is particularly interested in performance opportunities from mixed-precision algorithms for multiphysics simulations on modern computer architectures. 

Salman Habib is the Director of the Computational Science Division at Argonne National Laboratory with joint positions at The University of Chicago and Northwestern University. His research interests cover a wide range of problems in physics, ranging from cosmology to quantum information, with a major interest in supercomputing applications and algorithms. Habib leads the ExaSky project within the ECP. 

Christoph Junghans is the Deputy Group Leader of the applied computer science group at Los Alamos National Laboratory. 
His research interests span from scientific software development and engineering over molecular dynamics methods to multi-scale simulation techniques.

Damien Lebrun-Grandi\'e is a Computational Scientist at Oak Ridge National Laboratory.  He is co-maintainer of the Kokkos core library which provides performance portability to hundreds of scientific HPC applications, as well as the lead developer of the ArborX geometric search library.  Within CoPA, Damien is primarily involved with the development of Cabana.

Jamaludin Mohd-Yusof is a Scientist at Los Alamos National Laboratory. His interests include materials science, machine learning and fluid mechanics, where he develops novel algorithms and applications for High Performance Computing and emerging architectures. Within CoPA he primarily contributes to the BML effort. 

Stan Moore is a Staff Member at Sandia National Laboratories. He specializes in using Kokkos to extend particle-based simulation methods such as molecular dynamics to run efficiently on HPC platforms, and running particle-based simulations at large-scale. He is a core software developer of the LAMMPS molecular dynamics code.

Daniel Osei-Kuffuor is a Staff Scientist in the Center for Applied Scientific Computing (CASC) at Lawrence Livermore National Laboratory. His research interests include numerical linear algebra, sparse matrix computations, and scalable numerical solver and algorithm development for HPC applications, including electronic structure calculations. His work on CoPA supports the BML and PROGRESS libraries.

Steve Plimpton is a Staff Member at Sandia National Laboratories.  He has worked on a variety of particle-based methods and open-source simulation software, mostly for materials modeling.  He is a developer for the LAMMPS molecular dynamics package.

Adrian Pope is a Staff Scientist at Argonne National Laboratory. His research focuses on cosmological n-body simulations and statistical inference from astronomical surveys. He is a core developer of the HACC cosmological simulation code, maintains the stand-alone version of HACC's 3D FFT called SWFFT, and works with CoPA on potential technology transfer from HACC to other particle-based codes and solvers.

Samuel Temple Reeve is a Computational Scientist at Oak Ridge National Laboratory, formerly a postdoctoral researcher at Lawrence Livermore National Laboratory, working on the Cabana library and CabanaMD proxy app. His research interests span atomistic and microstructural simulation methods for problems in materials science.

Lee Ricketson is a Staff Scientist at Lawrence Livermore National Laboratory.  His research focuses on numerical methods for the kinetic equations governing plasma dynamics.  He is particularly interested in the advancement of particle-in-cell methods.

Aaron Scheinberg is a Computational Scientist focusing on exascale computing, scientific application performance, particle-based methods, magnetic fusion simulations, and GPU programming. Formerly at the Princeton Plasma Physics Laboratory, he is now a consultant at Jubilee Development.

Amil Y. Sharma is an Associate Research Physicist at the Princeton Plasma Physics Laboratory. He is a developer of the magnetic fusion simulation code XGC, which is part of the ECP WDMApp project.

Michael Wall is a Scientist at Los Alamos National Laboratory. His main expertise is in data processing and simulations for macromolecular X-ray diffraction studies. His recent focus has been on molecular-dynamics simulations for protein crystallography, parallel processing of diffuse X-ray scattering data, and quantum molecular-dynamics simulations of proteins. 

\end{biogs}

\end{document}